\documentclass[12pt]{article}

\pdfoutput=1

\usepackage[top=80pt,bottom=85pt,left=85pt,right=85pt]{geometry}
\usepackage{amssymb}
\usepackage{amsmath}
\usepackage{graphicx,color,subfigure}
\usepackage{float}
\usepackage{cite}
\usepackage[debug,pageanchor=false]{hyperref}
\definecolor{link}{rgb}{.8,.15,.1}
\hypersetup{colorlinks=true,linkcolor=link,citecolor=link,urlcolor=link,linktocpage}

\newcommand{\R}{\mathbb{R}}

\newcommand{\del}{\partial}

\makeatletter
\@addtoreset{equation}{section}
\makeatother

\newcommand{\beq}{\begin{equation}}
\newcommand{\eeq}{\end{equation}}
\newcommand{\bea}{\begin{eqnarray}}
\newcommand{\eea}{\end{eqnarray}}
\newcommand{\nn}{\nonumber}

\begin{document}

\begin{titlepage}

\begin{center}

\vskip .5in %.3in 
\noindent

{\Large \bf{Minimal flux Minkowski classification}}

\bigskip\medskip

Niall T. Macpherson and Alessandro Tomasiello\\

\bigskip\medskip
{\small 

Dipartimento di Fisica, Universit\`a di Milano--Bicocca, \\ Piazza della Scienza 3, I-20126 Milano, Italy \\ and \\ INFN, sezione di Milano--Bicocca
	
}

\vskip .5cm %.3cm
{\small \tt niall.macpherson@mib.infn.it, alessandro.tomasiello@unimib.it}
\vskip .9cm %.6cm
     	{\bf Abstract }

\vskip .1in
\end{center}

\noindent
We classify Minkowski$_4$ solutions in type IIA supergravity, with ${\cal N}=2$ supersymmetry and an SU(2) R-symmetry of a certain type. Many subcases can be reduced to relatively simple PDEs, among which we recover various intersecting brane systems, and AdS$_d$ solutions, $d=5,6,7$, and in particular the recently found general massive AdS$_7$ solutions. Imposing compactness of the internal six-manifold we obtain promising solutions with localized D-branes and O-planes.

\noindent

\vfill
\eject

\end{titlepage}

\tableofcontents

\section{Introduction} % (fold)
\label{sec:intro}

String theory compactifications become harder to find as the cosmological constant of the spacetime factor increases. There are many known families of AdS examples ($\Lambda<0$); the oldest known type is the Freund--Rubin class \cite{freund-rubin} which has a Sasaki--Einstein as internal space, but many other possibilities have been found over the years. 

Finding de Sitter examples ($\Lambda>0$) is infamously much harder. Several no-go results \cite{gibbons-nogo,dewit-smit-haridass,maldacena-nunez} imply that this is only possible at the cost of including quantum corrections and/or orientifold sources. Supersymmetry has to be broken in this case, which is of course to be expected at some scale anyway, but which makes finding such solutions harder. 

For Minkowski solutions $(\Lambda=0)$, supersymmetry can still be preserved. One expects intuitively this case to be much rarer than AdS already because setting $\Lambda=0$ looks like fine-tuning. Indeed the de Sitter no-go theorems in supergravity \cite{gibbons-nogo,dewit-smit-haridass,maldacena-nunez} also apply here, with the only exception of solutions where the metric is the only non-zero field. In this case the internal space has to be Ricci-flat. For compactifications of type II supergravity to four dimensions, these are Calabi--Yau manifolds. These exist in large numbers, and their study has been very fruitful to string theory and to geometry. But a more general study of Minkowski solutions can be useful both as a laboratory for string theory dynamics and as intermediate construction for the $\Lambda>0$ case.

To add the other supergravity fields, the antisymmetric fields often called ``fluxes'', one obvious possibility is to add D-branes: a probe analysis shows that these can be wrapped on complex or special Lagrangian cycles of the Calabi--Yau while preserving half of the supercharges. The Bianchi identity shows that one also has to add O-planes, in agreement with the general no-go theorems. One expects that the backreaction of these objects should then distort the internal metric. In general these metrics, while expected to exist, are not easy to exhibit explicitly (after all the Calabi--Yau metrics themselves are not known). For the special case of D3--O3 configurations, the distortion treats all the internal coordinates in the same way, and one ends up with a conformally Calabi--Yau $M_6$. Remarkably, one can also add a combination of three-form fluxes (as long as they are $(2,1)$ and primitive) without breaking supersymmetry \cite{grana-polchinski,giddings-kachru-polchinski,becker2,dasgupta-rajesh-sethi}. However, these solutions are still based on modifications of the Calabi--Yau geometry. 

In full generality, supersymmetry imposes \cite{gmpt2} that the internal $M_6$ should be a so-called ``generalized Calabi--Yau'' \cite{hitchin-gcy,gualtieri} with extra conditions involving the fluxes. Still, explicit examples where all these conditions are met are hard to obtain. Since several generalized Calabi--Yau's have been found, it is natural to start from those and to try to impose the extra flux conditions on them. One can obtain a few formal solutions in this way \cite{gmpt3}, but they rely on the presence of sources which are smeared all over $M_6$: this can be fine for D-branes, but not for O-planes, which are not dynamical and should sit at fixed loci of involutions. One can hope that such formal solutions are an ``approximation'' to more sensible ones where the sources are localized, but this needs to be demonstrated case by case.\footnote{One finds similar issues for AdS solutions, if one tries to make $\Lambda$ much smaller than the typical size of the internal space. This can be achieved in formal solutions with smeared O-planes \cite{acharya-benini-valandro,dewolfe-giryavets-kachru-taylor,petrini-solard-vanriet}. The localization of the O-planes of \cite{acharya-benini-valandro,dewolfe-giryavets-kachru-taylor} is in principle possible \cite{saracco-t}, but for the whole solution~ it remains to be demonstrated. It would be interesting to apply our present methods to this problem as well.} 

More sophisticated attempt have been made. For example \cite{andriot} looked for solutions with localized sources on solvmanifolds in bigger generality than in \cite{gmpt3}, finding examples again when the internal space is Ricci-flat. \cite{andriot-blaback-vanriet} recently looked at an Ansatz inspired by the conformal Calabi--Yau class \cite{grana-polchinski,giddings-kachru-polchinski,becker2,dasgupta-rajesh-sethi} and proved some general results.  Impressively, \cite{candelas-constantin-damian-larfors-morales-1,candelas-constantin-damian-larfors-morales-2} used a mix of generalized complex geometric and algebraic geometric methods to produce U-fold Minkowski$_4$ solutions. 
 
In this paper we take a different approach. Rather than making an Ansatz on the type of internal geometry or topology, we choose a broad class, and we let supersymmetry fix the internal geometry. This can succeed because, in some situations, there are enough internal spinors so that their bilinears can define an entire vielbein (in this context, an ``identity structure''). While this phenomenon has been known for a long time, it has become clearer in recent times that the local metrics thus naturally chosen by supersymmetry often tend to have automatically the correct behaviors one would expect from O-planes and D-branes. For example, AdS$_7$ solutions naturally allow for localized O6-sources \cite{afrt}, and the same is true for their AdS$_4$ cousins obtained by twisted compactification \cite{rota-t}. Since, as we mentioned, O-planes are essential for Minkowski compactifications, it feels natural to apply to them as well this ``identity structure'' approach. 

We thus analyze systematically a broad class of solutions, defined by its type of supersymmetry rather than by the internal topology. The class we chose lies at the intersection of several already existing physical constructions. It is rich enough for the classification to be interesting, and yet constrained enough for it to be very detailed. It consists of solutions with ${\cal N}=2$ unbroken supersymmetry, which admit an SU(2) R-symmetry acting in the simplest way, namely on an $S^2$ factor. We assumed the presence of the R-symmetry so as to allow us to recover various AdS$_d$ solutions with $d=5,6,7$; in particular the AdS$_7$ solutions recently obtained in massive IIA \cite{afrt,10letter} and the M-theory AdS$_5$ solutions of \cite{lin-lunin-maldacena}. (This latter point suggests that our results should be useful to look for ${\cal N}=2$ solutions in massive IIA as well.) 

Within our class, we reduce the classification to three broad subclasses (see figure \ref{fig:sum} below for a visual summary). Within each, preserved supersymmetry can be reduced to a system of PDEs; sometimes further sub-subclasses suggest themselves, and within each the PDEs simplify considerably. Some of these systems have appeared earlier in the literature for intersecting brane solutions \cite{youm,imamura-D8,janssen-meessen-ortin}. In fact the local form of the metric formally resembles in many cases that of a brane or of a brane system. One of the three subclasses is T-dual in its entirety to the conformal Calabi--Yau solutions \cite{grana-polchinski,giddings-kachru-polchinski,becker2,dasgupta-rajesh-sethi}. 

We will also illustrate the classification with some examples. As we already mentioned, we recovered several classes of AdS$_d$ solutions, $d=5,6,7$. We also studied the case where $M_6$ is compact. Even for the PDE systems that were already postulated in the literature for intersecting branes, this had not been really attempted. We will find a couple of workable examples; in particular one with an O6 and an ONS5. More work is required for a  more thorough understanding of the various possibilities offered by our results.  

In section \ref{sec:ansatz} we introduce our Ansatz. After some preliminary work in section \ref{sec:pure} about the pure spinor formalism we use, we give our detailed classification in section \ref{sec:class}, with a summary in section \ref{sub:sum}. We conclude with examples in section \ref{sec:ex}.

\textbf{Note added.} While this work was being completed, \cite{bobev-dibitetto-gautason-trujien} appeared, which has some overlap with our section \ref{sub:ads7} about recovering AdS$_7$ solutions.

% section intro (end)

\section{The Ansatz} % (fold)
\label{sec:ansatz}

In this section, we will motivate and introduce our Ansatz. We are looking for ${\cal N}=2$ Mink$_4$ solutions with an SU(2) R-symmetry acting on an $S^2$ factor.

As is customary we take the metric to be a warped product
\begin{equation}\label{eq:ds}
	ds^2_{10}= e^{2A} ds^2_{{\rm Mink}_4} + ds^2_6
\end{equation}
with the warping function $A$, the dilaton $\phi$ and the NS 3-form $H$  depending on the internal space $M_6$ only. 
The RR fluxes decompose as
\beq
F_{\rm tot}= \sum_{k=0}^5 F_{2k}= f + e^{4A}\text{Vol}_4\wedge \star_6\lambda(f),
\eeq
where $f$ denotes the total internal RR flux. The function $\lambda$ acts on an arbitrary $n$-form as $\lambda(\alpha_n)= (-1)^{\frac{n}{2}(n-1)}\alpha_n$ which ensures the higher and lower forms are related as $F_n = (-1)^{\frac{n}{2}(n-1)}\star_{10} F_{10-n}$. The Bianchi identities and equations of motion for the RR fluxes collectively read, away from any localised sources:
\beq
d_{H}F_{\rm tot} =0 \, ,\qquad d_{H} = d-H\wedge\,.
\eeq
The $d_H \star_6 \lambda(f)$ part of this follows from supersymmetry, while the $d_H f$ part needs to be imposed. The equation of motion for the three-form $H$ was also proven \cite{koerber-tsimpis} to follow from the supersymmetry equations, so we will not write it here.

The supersymmetry parameters for an ${\cal N}=2$ solution with flux in general read
\begin{equation}\label{eq:n2eps}
	\begin{split}
		&\epsilon_1 = \sum_{a=1}^2 \zeta^a_+\otimes \chi_{1+}^a + \text{c.c.}\\
		&\epsilon_2 = \sum_{a=1}^2 \zeta^a_+\otimes \chi_{2-}^a + \text{c.c.}
	\end{split}
\end{equation} 
where $\zeta$ and $\chi$ denote spinors on Mink$_4$ and $M_6$ respectively. (We define $(\zeta_+)^c = \zeta_-$, $(\chi_{b+}^a)^c = \chi_{b-}^a$, $b=1,2$.) We make the reasonable simplifying assumption that the spinors on $M_6$ have equal norm. This is a global requirement for the existence of AdS$_d$ solutions in $d=4,5,6,7$ and a local one for the existence of D-brane or O-plane sources, but is not needed in full generality.

Minkowski solutions don't necessarily have an R-symmetry. Eventually, however, we would like to apply our results to AdS$_d$ solutions with $d>4$. Indeed for example AdS$_5$ solutions can be viewed as Mink$_4$ solutions by taking in (\ref{eq:ds})
\begin{equation}\label{eq:mink-ads}
	e^A=e^{A_5 + \rho} \ ,\qquad  ds^2_6= e^{2A_5}d\rho^2 + ds^2_5\ .
\end{equation}
For AdS solutions R-symmetry has to be present: this is ultimately because it is necessarily a part of the superconformal algebra, even though it is optional in supersymmetry. In particular, AdS solutions with eight supercharges have an SU(2) R-symmetry. 

Thus we will assume the existence of an SU(2) R-symmetry. It will act on $M_6$ as an isometry. We will assume that it acts on $S^2$ leaves of a foliation on $M_6$. In other words, the metric will locally factorize as $S^2\times M_4$, with the radius of the $S^2$ depending on the coordinates on $M_4$ (including the possibility of shrinking to zero at some loci). This is what we will call the ``minimal'' class. It is not a priori the most general situation: one might take $M_4$ to be topologically fibred over $S^2$, for example with $M_6= M_3 \times$ a squashed $S^3$.

In this situation, the spinor Ansatz needs to be refined from the general (\ref{eq:n2eps}). The  $\zeta^a$ are a doublet under SU(2)$_{\rm R}$; thus we also need a doublet of spinors $\xi^a$ in the internal space, so that the ten-dimensional $\epsilon_{1,2}$ are invariant. In fact, since SU(2)$_{\rm R}$ acts only on the $S^2$, the $\xi^a$ need to transform as a doublet under the isometry group of $S^2$. Such a doublet is given by the so-called Killing spinors on $S^2$, which is also related to its Killing vectors. It is of the form 
\begin{equation}
	\xi^a = \left(\begin{array}{c} \xi \\ \xi^c
	\end{array}\right)\ ;
\end{equation}
see appendix \ref{app:s2} for details. 

We can also play with chirality projections, to obtain doublets $\zeta^a_\pm$ and $\xi^a_\pm$ on Mink$_4$ and $S^2$ respectively. So there are four possible SU(2)$_{\rm R}$ singlets; this leads us to\footnote{The class we selected in this paper was also partially inspired by the AdS$_4$ solutions in \cite{rota-t}; in that paper, however, the doublet on $S^2$ is paired up with a doublet on a factor $\Sigma_3$ of the internal $M_6$. In that case the R-symmetry is absent in the AdS$_4$ solution, because the solutions are ${\cal N}=1$.}  
\begin{equation}\label{eq:geneps}
	\begin{split}
	\epsilon_1&= \sum_{a=1}^2 \zeta^a_{+}\otimes\xi^a_{+}\otimes\eta^1_++\zeta^a_{+}\otimes\xi^a_{-}\otimes\eta^1_-+\zeta^a_{-}\otimes \xi^a_{-}\otimes \tilde{\eta}^1_++\zeta^a_{-}\otimes \xi^a_{+}\otimes \tilde{\eta}^1_-+ \text{c.c.},\\
	\epsilon_2&=\sum_{a=1}^2  \zeta^a_{+}\otimes\xi^a_{+}\otimes\eta^2_-+\zeta^a_{+}\otimes\xi^a_{-}\otimes\eta^2_++\zeta^a_{-}\otimes \xi^a_{-}\otimes \tilde{\eta}^2_-+\zeta^a_{-}\otimes \xi^a_{+}\otimes \tilde{\eta}^2_++ \text{c.c},
	\end{split}
\end{equation}
where the $\eta^a$, $\tilde \eta^a$ are now spinors on $M_4$. When one writes out the supersymmetry equations, however, one actually realizes the $\eta$'s and $\tilde\eta$'s never mix.\footnote{This can roughly be seen like this. First, as is standard, one can separate the part of each equation multiplying $\zeta^a_+$ from the part multiplying $\zeta^a_-$, to obtain two conjugate equations on $M_6$. Focusing on the positive chirality one, one can then further separate this into parts multiplying $\xi^a_+$, $\xi^a_-$ and their conjugates. Crucially, these are all independent: their functional dependences are all different, even if at every point on $S^2$ there are only two spinors. Now one can check that the equation arising from the term multiplying $\xi_+$ only contains $\eta$'s, the equation arising from $\xi^c_-$ only contains $\tilde \eta$'s, and so on. This would not be true for AdS$_4$ solutions, or if there were any any fluxes with only one leg  along the  $S^2$; the latter are forbidden in our situation by the SU(2)$_{\rm R}$ symmetry.} Thus including the $\tilde \eta$'s only gives more constraints, beyond those that are necessary for minimal supersymmetry in four dimensions, and not a generalization; we can then set them to zero without loss of generality. So we are left with  
\begin{equation}\label{eq:mineps}
	\begin{split}
	\epsilon_1&=\sum_{a=1}^2 \zeta^a_{+}\otimes\xi^a_{+}\otimes\eta^1_++\zeta^a_{+}\otimes\xi^a_{-}\otimes\eta^1_-+ \text{c.c.},\\
	\epsilon_2&=\sum_{a=1}^2 \zeta^a_{+}\otimes\xi^a_{+}\otimes\eta^2_-+\zeta^a_{+}\otimes\xi^a_{-}\otimes\eta^2_++ \text{c.c}.
	\end{split}
\end{equation}
One of the first condition one encounters when working with (\ref{eq:geneps}) is that the norms of the $\chi_b^a$ should be proportional to $e^A$ \cite[App.~A.3]{gmpt3}. 
For SU(2)$_{\rm R}$ to be unbroken, one needs this to be independent of the coordinates on $S^2$; which implies
\begin{equation}\label{eq:zeroconsimp}
	|| \eta^a_+ ||^2 = || \eta^a_- ||^2\ .
\end{equation}

In the rest of the paper, we will classify ``minimal'' solutions using (\ref{eq:mineps}) as a starting point. As we will see, the geometry is already constrained enough that a very detailed classification is possible; this is ultimately due to the fact that the internal spinors $\chi^a_b$ define an identity structure. 

% section ansatz (end)

\section{Pure spinors} % (fold)
\label{sec:pure}

Having described our ``minimal'' class of manifolds, we will now classify its solutions. We will use the pure spinor formalism \cite{gmpt2}, with our precise conversions .

First of all, there is no need for us to consider all the $\chi^a_b$. If we impose that an ${\cal N}=1$ subalgebra is preserved (corresponding for example to the $\chi_b \equiv \chi_b^1$), R-symmetry will imply automatically that the second set of supercharges is also preserved. So we can simply work with the two internal spinors 
\begin{equation}\label{eq: simp6dspinors}
\begin{split}
	\chi^1_+&=e^{\frac{A}{2}}(\xi_+\otimes\eta^1_++\xi_-\otimes\eta^1_-)%=\frac{1}{2}\left(\xi\otimes \eta^1+\sigma_3\xi\otimes \hat \gamma\eta^1 \right)\ 
	,\\
	\chi^2_-&=e^{\frac{A}{2}}(\xi_+\otimes\eta^2_-+\xi_-\otimes\eta^2_+)
	%=\frac{1}{2}\left(\xi\otimes \eta^2-\sigma_3\xi\otimes \hat \gamma\eta^2\right)
	\ .	
\end{split}
\end{equation}
If we define the bispinors 
\beq
\Phi_- = e^{-A}\chi^1_+\otimes \chi^{2\dag}_-\ , \qquad \Phi_+=e^{-A}\chi^1_+\otimes \overline{\chi}^{2}_-,
\eeq
where $\overline{\chi}= (\chi^c)^{\dag}$, the differential forms associated to them via the Clifford map are pure spinors for the Cl$(6,6)$ algebra living on the ``generalized tangent bundle'' $T\oplus T^*$. Preserved supersymmetry is equivalent to the following ``pure spinor equations'' \cite{gmpt2} on $M_6$: 
\begin{subequations}\label{eq:psp6d}
	\begin{align}
	d_{H}\left(e^{3A-\Phi}\Phi_+\right)&=0, \label{eq:Phi+}\\
	d_{H}\left(e^{2A-\Phi}\text{Re}\Phi_-\right)&=0, \label{eq:RePhi-}\\
	d_{H}\left(e^{4A-\Phi}\text{Im}\Phi_-\right)&=\frac{e^{4A}}{8}\star_6\lambda(f),
	\label{eq:ImPhi-}
	\end{align}	
\end{subequations}
where $\lambda(f)=\lambda(f_0 + f_2 + f_4 + f_6)= f_0 - f_2 +f_4 -f_6$.

The decomposition (\ref{eq: simp6dspinors}) induces a decomposition of the pure spinors $\Phi_\pm$; this can be used to reduce the system (\ref{eq:psp6d}) to one on $M_4$. One can define
\begin{equation}\label{eq:nochiralPSI}
	\begin{split}
	&\Psi= \eta^1\otimes\eta^{2\dag}\ ,\qquad \tilde{\Psi}= \eta^1\otimes\overline{\eta}^{2},\\
	&\Psi_{\hat{\gamma}}= (\hat{\gamma}\eta^1)\otimes\eta^{2\dag}\ ,\qquad\tilde{\Psi}_{\hat{\gamma}}= (\hat{\gamma}\eta^1)\otimes\overline{\eta}^{2}\ ,.
	\end{split}
\end{equation}
where $\eta^i= \eta^i_+ + \eta^i_-$ and $\hat \gamma$ is the chiral gamma in $M_4$. It is also convenient to split metric and fluxes as 
\begin{subequations}
\begin{align}
ds^2_6=&e^{2C}ds^2(S^2)+ ds^2(M_4),\\[2mm]
\label{eq:Bdec} B=& B_2+e^{2C} B_0 \text{Vol}(S^2),\\[2mm]
f=& F+e^{2C} G\wedge \text{Vol}(S^2)\ .
\end{align}
\end{subequations}
after defining functions $C$, $B_0$ and forms $B_2$, $F$, $G$ on $M_4$. One obtains in this way several equations for $\Psi$, $\Psi_{\hat \gamma}$, $\tilde{\Psi}$ and $\tilde{\Psi}_{\hat \gamma}$, which we give in appendix \ref{app:psp4}. One of the consequences is that the zero-form parts of both $\Psi$ and $\tilde{\Psi}_{\hat \gamma}$ must vanish:
\begin{equation}\label{eq:0f}
	\Psi_0 = (\tilde\Psi_{\hat \gamma})_0 = 0 \ .
\end{equation}

A spinor $\eta$ in four dimensions defines an associated basis
\begin{equation}\label{eq:eta-basis}
	\{ \eta, \eta^c, \hat \gamma \eta, \hat \gamma \eta^c\}
\end{equation}
for the space of spinors. Thus we can expand $\eta^2$ on the basis associated to $\eta^1$:
\begin{equation}\label{eq:eta12par}
	\eta^1= \eta \ ,\qquad \eta^2= a_0\eta+a \hat{\gamma}\eta+ b \eta^c+b_0 \hat{\gamma}\eta^c\ .
\end{equation}
\eqref{eq:zeroconsimp} makes (\ref{eq:eta-basis}) orthogonal and $\eta$ have constant norm, and since $|\eta^1|=|\eta^2|$, it follows that  $|a_0|^2+|b_0|^2+|a|^2+|b|^2=1$. However since $\eta^{2\dag}\eta^1= \overline{a}_0$ and $\overline{\eta}^{2}\hat{\gamma}\eta^1= \overline{b}_0$, we must set $a_0=b_0=0$ to satisfy (\ref{eq:0f}); so we have
\begin{equation}\label{eq:ab}
	|a|^2+|b|^2=1\ .
\end{equation}
One expects from (\ref{eq:0f}) an orthogonal SU(2) structure on $M_6$. We can parameterize it as in \cite[Sec.~3.2]{afprt} using the spinors  to define the vielbein
\begin{equation}\label{eq:vw}
	v_m = \eta^{\dagger}_- \gamma_m \eta_+ \ ,\qquad
	w_m = \overline{\eta_-} \gamma_m \eta_+\ .
\end{equation}
The pure spinors $\Phi_\pm$ on $M_6$ then read
\beq \label{eq:pspE}
\Phi_+ =\frac{1}{8} E_1\wedge E_2 \wedge e^{\frac{1}{2} E_3\wedge \overline{E}_3}\ ,\qquad\Phi_-=\frac{1}{8} E_3\wedge e^{\frac{1}{2}(E_1\wedge \overline{E}_1+E_2\wedge \overline{E}_2)}
\eeq
where
\begin{align}\label{eq:canonical6d vielbein}
E_1&= b (e^{C} dy_3- y_3 v_2)+i (a w + b v_1),\nn\\[2mm]
E_2&= -(e^C d( y_1+i y_2)- ( y_1+i y_2) v_2),\\[2mm]
E_3&= i \overline{a} (e^{C} dy_3- y_3 v_2)+(\overline{b} w- \overline{a} v_1) \ .\nn
\end{align}
Here $v=v_1+i v_2$, $w=w_1+i w_2$, and the $y_a$ are the three ``embedding coordinates'' of $S^2$ defined in (\ref{eq:ya}).

In terms of the parameterization in (\ref{eq:eta12par}), with  $a_0=b_0=0$, the pure spinors on $M_4$ can be written as
\begin{subequations}\label{4d bi}
	\begin{align}
	\Psi^{\text{odd}}&=-\frac{1}{2}i \overline{a} v_2\wedge e^{\frac{1}{2}w\wedge \overline{w}-\frac{\overline{b}}{\overline{a}}v_1\wedge w}\ ,\qquad\Psi^{\text{even}}=-\frac{1}{2}iv_2\wedge(\overline{b}w- \overline{a} v_1)\wedge e^{\frac{1}{2}w\wedge \overline{w}},\\[2mm]
	\Psi_{\hat\gamma}^{\text{odd}}&= \frac{1}{2}\big(\overline{b}w-\overline{a} v_1 \big)\wedge e^{\frac{1}{2}w\wedge \overline{w}}\ ,\qquad\Psi_{\hat\gamma}^{\text{even}}= \frac{1}{2}\overline{a} e^{\frac{1}{2}w\wedge \overline{w}-\frac{\overline{b}}{\overline{a}} v_1\wedge w},\\[2mm]
	\tilde{\Psi}^{\text{odd}}&= -\frac{1}{2}\big(a w+ b v_1\big)\wedge e^{\frac{1}{2}w\wedge \overline{w}}\ ,\qquad\tilde{\Psi}^{\text{even}}=-\frac{1}{4}b e^{\frac{1}{2}w\wedge \overline{w}+\frac{a}{b} v_1\wedge w},\\[2mm]
	\tilde{\Psi}_{\hat\gamma}^{\text{odd}}&=-\frac{1}{2}i bv_2 \wedge e^{\frac{1}{2}w\wedge \overline{w}+\frac{a}{b}v_1\wedge w}\ ,\qquad\tilde{\Psi}_{\hat\gamma}^{\text{even}}=-\frac{1}{2}i v_2\wedge(a w+ b v_1)\wedge e^{\frac{1}{2}w\wedge \overline{w}}.
	\end{align}
\end{subequations}
We can actually take $b$ to be real. To see this, notice that (\ref{4d bi}) and \eqref{eq: Phip b} yield the 1-form constraint
\beq
d( e^{3A+2C-\Phi}b)+ 2 b e^{3A+2C-\Phi} v_2=0.
\eeq
This implies that $b d\overline{b}=\overline{b} db$ and so $b= |b|e^{i\beta_0}$ for $d\beta_0=0$. Now, sending $w\to e^{i \beta_0}w$ in \eqref{4d bi} is equivalent to sending $b,\overline{b}\to|b|$ everywhere they appear and multiplying $\tilde{\Psi}$ and $\tilde{\Psi}_{\hat\gamma}$ by $e^{i\beta_0}$ in \eqref{eq: Phip a}--\eqref{eq: Phip b}. However, since $\beta_0$ is constant, these phases leave these conditions unchanged; so without loss of generality we can set $\beta_0=0$. We now expand $a\equiv a_1+i a_2$, so that (\ref{eq:ab}) now reads
\beq \label{eq:a12b}
a_1^2+a_2^2+b^2=1\ .
\eeq
Notice that the analysis in this part of the paper is very similar to the AdS$_5$ classification in \cite{afpt}.  In particular, we are getting an orthogonal SU(2) structure, just like in that case. It would be interesting to explore to what extent this gets generalized if one abandons our assumption, made in section \ref{sec:ansatz} that the metric is locally an $S^2 \times M_4$ product, by topologically fibering part of the $M_4$ over the $S^2$. 

Since we have $a^{-1}$ and $b^{-1}$ appearing in \eqref{4d bi}, we should treat the cases where these vanish individually, before performing a general analysis for arbitrary non zero $a$ and $b$. This is what we will now proceed to do.

% section pure (end)

\section{Classification} % (fold)
\label{sec:class}

We have obtained a pure spinor parameterization (\ref{4d bi}) in terms of three real functions $a_1$, $a_2$, $b$. Some of the cases when one or more of these parameters vanish have to be studied separately. This gives rise to a ramified structure; we will study it in detail in this section. Sections \ref{sub:6dsols} and \ref{sub:5dsols} are not strictly needed in the ramification, in the sense that their results can be obtained as limits from sections \ref{sub:generic} and \ref{sub:a10case} respectively. Sections \ref{sub:4dsols}, \ref{sub:a10case} and \ref{sub:generic} are instead all substantially different, and have to be dealt with separately. We give a brief summary in section \ref{sub:sum}.

Before we begin, let us make some general comments about our methods. Using the supersymmetry equations, we get local expressions for the metric and the fluxes, as well as some PDEs. We then impose the Bianchi identities to be satisfied almost everywhere: this yields some additional PDEs. When we interpret the resulting solutions physically, we often find that they in fact involve one or more localized object, such as a D-brane or an orientifold; not surprisingly, this is always the case when the internal space $M_6$ is compact. One then needs to check, in each example, that the Bianchi identities are not only valid away from these localized sources, but also on them, with the appropriate delta-like source term included. Practically speaking, this is best done (just like in electromagnetism) by checking the integral version of the Bianchi identities. In fact, the behaviour of the metric and of the other fields near a localized object is a good guide to which delta-like terms are present; this is because the local behaviour is in fact usually locally identical (via a change of coordinates) to that of a brane or a brane system in flat space.  In this section, we will find the local form of the fields and the PDEs to be solved. A detailed treatment of the sources needs to be done on a case-by-case basis, and we will do so for some explicit examples in section \ref{sec:ex}. However, already in this section we will make comments about which sources we expect to be present in a given class, based on the fields that are present and on the structure of the metric.

\subsection{The $a_2=b=0$ case: $M_6=S^2\times T^2\times M_2$} % (fold)
\label{sub:6dsols}
We begin by examining the case where $b=a_2=0$, choosing $a_1=1$ to satisfy \eqref{eq:a12b} without loss of generality. In fact this case can be obtained from the ``generic'' case of section \ref{sub:generic}, in the sense that the solutions of the current subsection can be obtained by taking the $b\to 0$, $a_2\to 0$ limit from that subsection. However, we found it clearer to deal with this particular case separately in the present subsection, especially since the solutions obtained here can be used as seed solutions for the ``generic'' ones.

Upon inserting the definitions of \eqref{4d bi} into (\ref{eq: Phip})--(\ref{eq: ImPhim}) we are able to show that the conditions for unbroken supersymmetry reduce to
\begin{subequations}
\begin{align}
&B_2=0,\label{ba20 SUSY a}\\[2mm]
&d(e^{-A}w)=0,\qquad d(e^{2A-\Phi}v_1)=0,\qquad d(e^{4A+C-\Phi})+ e^{4A-\Phi}v_2=0,\label{ba20 SUSY b}\\[2mm]
&d(e^{-4A+\Phi}(v_1+ B_0 v_2))=0,\qquad d(e^{2A+2C-\Phi} (B_0 v_1-v_2))=0\label{ba20 SUSY c}.
\end{align}
\end{subequations}
This dramatic simplification of the supersymmetry conditions, and the techniques we will use to solve them, are prototypical of what we find for the entire minimal class we study, as such we will give more details here than we shall in subsequent sections. 

(\ref{ba20 SUSY a}) means that the NS three-form must have two of its legs on $S^2$. The next conditions (\ref{ba20 SUSY b}) can be solved by defining local coordinates $x_1,x_3,x_4$ and
\beq
x_2= e^{4A+C-\Phi},
\eeq
in terms of which the vielbein on $M_4$ is completely determined as
\beq\label{vielbein ba20}
v_1=- e^{-2A+\Phi}dx_1,~~v_2=-e^{-4A+\Phi}dx_2,\qquad  w= e^{A}(dx_3+i dx_4).
\eeq
The remaining conditions (\ref{ba20 SUSY c}) then give rise to the first order PDEs
\begin{subequations} \label{eq: BPS ba20}
\begin{align}
&\partial_{x_2}(e^{-6A+2\Phi})+ \partial_{x_1}(e^{-8A+2\Phi}B_0)=0,\label{eq: BPS ba20 a}\\[2mm]
&\partial_{x_2}(e^{-8A+2\Phi}B_0x_2^2)-\partial_{x_1}(e^{-10A+2\Phi}x_2^2)=0,\label{eq: BPS ba20 b}
\end{align}
\end{subequations}
and inform us that $e^{A},e^{C},B_0,\Phi$ depend on $x_1, x_2$ only. In other words, $\partial_{x_3}$ and $\partial_{x_4}$ are isometries of the metric which define a 2-torus, so that the internal manifold $M_6$ has a metric of the form
\beq\label{ba20 6dmet}
ds^2_6 = e^{2A}ds^2(T^2)+e^{-4A+2\Phi}dx_1^2+e^{-8A+2\Phi}\bigg(dx_2^2+ x_2^2 ds^2(S^2)\bigg).
\eeq
We see that $x_2$ is naturally a radial coordinate. Moreover, the metric is formally of the type one gets by superimposing a D6 whose harmonic function is $e^{-4A}$, and an NS5 whose harmonic function is $e^{-6A+2 \phi}$. The calibration for a spacetime-filling D-brane indeed indicates that a probe D6 can be wrapped along the $T^2$ and direction $x^1$. (We also see the possibility of a D8 transverse to $x^1$, compatibly with what we will see in section \ref{sub:imamass}.) The last parenthesis in (\ref{ba20 6dmet}) looks like an $\R^3$, but it can potentially be made compact with the help of the prefactor; we will come back to this point in section \ref{sub:comp}. 

The next thing we need to determine are the RR fluxes. These follow from (\ref{eq:4dflux}), which reduce in this case to
\begin{subequations}
\begin{align}
2e^{4A}\star_4 G&= i d(e^{4A-\Phi} v_1\wedge w\wedge \overline{w}),\\[2mm]
2e^{4A+2C}\star_4 F&=- i d(e^{4A-\Phi} v_2\wedge w\wedge \overline{w})+i e^{4A-\Phi} d(e^{2C}B_0)\wedge v_1\wedge w\wedge \overline{w}\ .
\end{align}
\end{subequations}
We can then use \eqref{vielbein ba20} to take the Hodge dual of these expressions and arrive at 
\begin{subequations}
\begin{align}
B&= e^{-8A+2\Phi}B_0 \text{Vol}(S^2),\label{eq: Fluxes ba20 a}\\[2mm]
F_0 &=2 e^{-2\Phi}\partial_{x_1}(e^{2A}),\label{eq: Fluxes ba20 b}\\[2mm]
F_2&=B F_0 -x_2^2\big(\partial_{x_2}(e^{-4A})-e^{-2A}B_0\partial_{x_1}(e^{-4A})\big)\text{Vol}(S^2)\label{eq: Fluxes ba20 c}.
\end{align}
\end{subequations}

The fluxes do not depend on the $T^2$ directions and they have no legs along them. Moreover, the factor in front of the $T^2$ metric in (\ref{ba20 6dmet}) is $e^{2A}$; so the metric can be reassembled as $ds^2_{\R^4\times T^2} + ds^2_4$. In the local analysis, the present case is then identical to $\rm{Mink}_6$ compactifications.

The last thing we need to impose to ensure these are supergravity solutions is that the Bianchi identities of the RR fluxes are satisfied. In this case this merely requires that $F_0$ is constant and that 
\beq\label{ba20 bianchi}
\partial_{x_2}(e^{-4A})-e^{-2A}B_0\partial_{x_1}(e^{-4A})=-\frac{c_1}{x_2^2}
\eeq
for some constant $c_1$. This in addition to \eqref{eq: BPS ba20 a}--\eqref{eq: BPS ba20 b} gives 3 PDE's that need to be solved. However, by stipulating whether we have the Romans mass turned on or not, we can reduce them to a single PDE for each case --- which we proceed to do.

\subsubsection{$F_0=0$ case}\label{sub:ima0}
From \eqref{eq: Fluxes ba20 b} it is clear that imposing $F_0=0$ requires that $A=A(x_2)$, from which it follows that
\beq \label{eq:e-4Aima}
e^{-4A}=c_2+ \frac{c_1}{x_2},
\eeq
which is behaviour appropriate for the warp factor of either a stack of D6-branes or an O6-plane, depending on the sign of $c_1$. Indeed if we also set $B_0=0$ we are quickly led to $e^{\Phi}=e^{3A}$, and  \eqref{ba20 6dmet} becomes that of the flat space D6/O6  metric. For $B_0\neq 0$ things are more complicated, but we can make progress by noting that \eqref{eq: BPS ba20 a} is an integrability condition which implies
\beq \label{eq:Adh}
e^{-6A+2\Phi}=\partial_{x_1}h(x_1,x_2),\qquad B_0 =e^{2A}\frac{\partial_{x_2}h(x_1,x_2)}{\partial_{x_1}h(x_1,x_2)}\ .
\eeq
From \eqref{eq: BPS ba20 b} we then get a single PDE to solve:
\beq\label{eq: h7dcond}
\frac{1}{x_2^2}\partial_{x_2}(x_2^2 \partial_{x_2} h)+ e^{-4A}\partial^2_{x_1}h=0\ .
\eeq
%Before moving on we summarise the data of the solutions we find in this subsection
%\begin{align}
%ds^2_6&=  e^{-2A} \partial_{x_1}h\big(e^{4A}dx_1^2+ dx_2^2+x_2^2ds^2(S^2)\big),\qquad F_2=c_1 \text{Vol}(S^2),\nn\\[2mm]
%B&= x_2^2 \partial_{x_2} h\text{Vol}(S^2),\qquad e^{-4A}=c_2+ \frac{c_1}{x_2},\qquad e^{-6A+2\Phi}=\partial_{x_1}h,
%\end{align}
%where $h$ is a solution to \eqref{eq: h7dcond}.
Notice that the first term is a Laplacian on $dx_2^2 + x_2^2 ds^2_{S^2}.$
This is a particular case of the system used in \cite{imamura-D8,janssen-meessen-ortin} to investigate NS5--D6 intersecting branes; this is related to our comment below (\ref{ba20 6dmet}).\footnote{\label{foot:ima}In the language of \cite{imamura-D8}, the metric is written as $ds^2=S^{-1/2}ds^2_{{\rm Mink}_6} + K (S^{-1/2} dz^2 + S^{1/2} ds^2_{\R^3})$; the supersymmetry and Bianchi equations reduce to $F_0 K= - 4\del_z S$, $\Delta K + \del_z^2(SK)=0$, $\Delta S + \frac12\del_z^2(S^2)=0$. For $F_0=0$, this is $\del_z S= \Delta S=0$, $\Delta K+ S\del_z^2 K=0$, which incidentally also looks similar to the equation in \cite{youm}. This corresponds to our case, identifying $x_1$ with $z$, $S=e^{-4A}$, $K=e^{-6A+2 \phi}$ and taking $\del_{x_1}$ of (\ref{eq: h7dcond}). In the $F_0\neq 0$ case, the equation for $K$ follows from the one for $S$ and can be dropped.}  

\subsubsection{$F_0\neq 0$ case}\label{sub:imamass}
With $F_0\neq 0$ we can use \eqref{eq: Fluxes ba20 b} to define the dilaton and \eqref{ba20 bianchi} to define $B_0$ as
\beq
B_0 =e^{2A}\frac{ c_1+ x_2^2 \partial_{x_2}(e^{-4A})}{x_2^2\partial_{x_1}(e^{-4A})},\qquad  e^{2\Phi}=\frac{2}{F_0 } \partial_{x_1}e^{2A}.
\eeq 
These definitions solve \eqref{eq: BPS ba20 a} automatically and reduce \eqref{eq: BPS ba20 b} to a PDE in $e^{-4A}$ only, namely
\beq\label{eq: PDEA}
\frac{1}{x_2^{2}}\partial_{x_2}(x_2^2\partial_{x_2}(e^{-4A}))+\frac{1}{2}\partial_{x_1}^2(e^{-8A})=0.
\eeq
Again this reduces to \cite{imamura-D8,janssen-meessen-ortin}  (see footnote \ref{foot:ima}), this time to be interpreted as a system relevant to NS5--D6--D8 brane systems.

% subsection 6dsols (end)

\subsection{The $a=0$ case: $M_6=S^2\times S^1\times M_3$} % (fold)
\label{sub:5dsols}
In this subsection we set $a=0$ and choose $b=1$ to satisfy \eqref{eq:a12b}. This case is actually a subcase of the case $a_1=0$, $b\neq 0$, which we will analyze later in section \ref{sub:a10case}, in the sense that the solutions of the present subsection can all be obtained by taking the $a_2\to 0$ limit in section \ref{sub:a10case}. In particular, the present subcase will turn out to be related to conformal Calabi--Yau solutions \cite{grana-polchinski,giddings-kachru-polchinski,becker2,dasgupta-rajesh-sethi}, since the larger case of section \ref{sub:a10case} will be. 
Nevertheless, we present the $a=0$, $b=1$ subcase separately for clarity.

The necessary and sufficient conditions for unbroken supersymmetry become
\begin{subequations}
\begin{align}
&B_2=-B_0 v_1\wedge v_2,\label{a1a20 SUSY a}\\[2mm]
&d(e^{A}\text{Re}w)=0,\qquad d(e^{-A}\text{Im}w)=0,\label{a1a20 SUSY b}\\[2mm]
&d(e^{3/2A-1/2\Phi} v_1)=0,\qquad d(e^{3A+2C-\Phi})+2 e^{3A+C-\Phi}v_2=0,\label{a1a20 SUSY c}\\[2mm]
&d(e^{2A-\Phi}\text{Re}w)=0,\qquad d(e^{2A+2C-\Phi}B_0\text{Re}w)=0,\label{a1a20 SUSY d}\\[2mm]
&d\big((1+B_0^2)e^{2C}v_1\wedge v_2 \big)\wedge \text{Re}w=0,\label{a1a20 SUSY e}.
\end{align}
\end{subequations}
The first thing we see is that \eqref{a1a20 SUSY a} implies that, unlike the case of section \ref{sub:6dsols}, the generic NS 3-form has a contribution orthogonal to the $S^2$ directions. As before we can solve several of supersymmetry conditions by appropriately choosing local coordinates that define the vielbein on $M_4$.
We take
\begin{align}\label{eq: azero vielbein}
v_1 &= e^{-3/2A+\frac{1}{2}\Phi}dx_1,\qquad  v_2= -e^{-3A-C+\Phi}x_2 dx_2,\nn\\[2mm]
 w&= -(e^{-A}dx_3+i e^{A}dx_4) \ ,\qquad x^2_2=e^{3A+2C-\Phi}\ .
\end{align}
This solves \eqref{a1a20 SUSY b}--\eqref{a1a20 SUSY c}.  The remaining conditions \eqref{a1a20 SUSY d}--\eqref{a1a20 SUSY e} are then uniquely solved by
\beq\label{eq:a1a20  BPS}
e^{A-\Phi}=f(x_3),\qquad e^{2C}B_0 = g(x_3),\qquad  \partial_{x_4}A=0,
\eeq
from which it follows that $\partial_{x_4}$ is an isometry. We take $\partial_{x_4}$ to define a $S^1$ so that the internal manifold $M_6$ has a metric of the form
\beq\label{a1a20 metric}
ds^2_6= e^{2A}ds^2(S^1) + \frac{e^{-2A}}{f}\bigg(dx_1^2+ dx_2^2+f dx_3^2+ x_2^2  ds^2(S^2)\bigg)\ .
\eeq
Formally this looks like a superposition of a D4 with harmonic function $e^{-4A}$ and of an NS5 with harmonic function $1/f$, although we will see that the interpretation is a bit more subtle. Similarly to (\ref{ba20 6dmet}), we can check that the calibration for a spacetime-filling D-brane indicates that a D4 can be BPS along the $S^1$ direction (A D8 on transverse to $x_3$ is another possibility). 

The fluxes are extracted as before by inserting \eqref{4d bi} into (\ref{eq:4dflux}) and then using \eqref{eq: azero vielbein} to take the Hodge dual. We finally arrive at
\begin{subequations}
\begin{align}
B&= g \,\mathcal{C}_2,\qquad ~\mathcal{C}_2= \frac{dx_1\wedge dx_2}{x_2^2}+ \text{Vol}(S^2) ,\label{eq: azero fluxes a}\\[2mm]
F_0& = \partial_{x_3}f,\qquad F_2= B F_0-\partial_{x_3}(fg)\mathcal{C}_2= -f\partial_{x_3}g\mathcal{C}_2,\label{eq: azero fluxes b}\\[2mm]
F_4& =B\wedge F_2-\frac{1}{2} B\wedge B F_0+\frac{1}{2}\partial_{x_3}(fg^2)\mathcal{C}_2\wedge \mathcal{C}_2\label{eq: azero fluxes c}\\[2mm]
&+x_2^2\bigg(\partial_{x_1}(e^{-4A})dx_2\wedge dx_3-\partial_{x_2}(e^{-4A})dx_1\wedge dx_3+\partial_{x_3}(f^{-1} e^{-4A})dx_1\wedge dx_2\bigg)\wedge\text{Vol}(S^2)\nn \ .
\end{align}
\end{subequations}
These have no legs on the $S^1$ spanned by $x_4$. So, analogously to what happened in section \ref{sub:6dsols}, the solution has an $\text{Mink}_4\times S^1$ factor, and can be locally viewed as a $\text{Mink}_5$ compactification.

Ensuring that the parts of \eqref{eq: azero fluxes b} that do not manifestly give rise to the Bianchi identities are closed then implies
\beq\label{eq: azero PDEs}
\partial^2_{x_3}f=0,\qquad ~ \partial^2_{x_3}(fg) = 0,
\eeq
which lead in general to
\beq\label{eq: azero PDEssol}
f= c_1+F_0 x_3,\qquad fg= (c_2+c_3x_3).
\eeq
On the other hand \eqref{eq: azero fluxes c} implies that the $F_4$ Bianchi identity follows from the PDE  
\beq\label{eq: a0 PDE}
\partial^2_{x_1}(e^{-4A})+\frac{1}{x_2^2}\partial_{x_2}(x_2^2\partial_{x_2}(e^{-4A}))+\partial^2_{x_3}(f^{-1} e^{-4A})+\frac{1}{x_2^4}\partial^2_{x_3}(f g^2)=0,
\eeq
which is the only thing left to solve. 

We notice this is a generalisation of the PDE leading to the fully localised D4-D8 system of \cite{youm}, reducing to it when $g=0$. Actually, more generally only
\begin{equation}\label{eq:dx3g}
	\del_{x_3} g= 0 \ 
\end{equation}
is required, as in this case the influence of $g$ is a pure NS 2-form. 

The final term in \eqref{eq: a0 PDE} makes it hard to solve in general. To make progress, we define $A$ in terms of an arbitrary function $h(x_1,x_2,x_3)$ as 
\beq
e^{-4A}= \frac{f}{ x_2^2} \bigg[x_2^2 h(x_1,x_2,x_3) - (\partial_{x_3}g)^2\bigg]
\eeq
from which it follows that
\beq
\partial^2_{x_3}h+f\bigg(\frac{1}{x_2^2}\partial_{x_2}(x_2^2\partial_{x_2}h)+\partial^2_{x_1}h\bigg)= 20\left(\frac{\partial_{x_3}f \partial_{x_3}g}{f x_2}\right)^2.
\eeq
This suggests an additional way to solve the PDE (\ref{eq: a0 PDE}), namely
\beq \label{eq:dx3f}
\partial_{x_3}f=0,
\eeq 
which is equivalent to $F_0=0$.

We will now analyze the cases (\ref{eq:dx3g}) and (\ref{eq:dx3f}) in turn. 

\subsubsection{$F_0=0$ Ansatz}
When $F_0=0$ we can without loss of generality set $f=1$ and the solutions are defined as
\begin{align} \label{eq:a0F0}
ds^2_6&= e^{2A}ds^2(S^1) + e^{-2A}\bigg(dx_1^2+ dx_2^2+ dx_3^2+ x_2^2  ds^2(S^2)\bigg),\qquad  e^{-4A}=e^{-4\Phi}= \frac{x_2^2 h - c^2}{ x_2^2} ,\nn\\[2mm]
B&= c x_3 \mathcal{C}_2,\qquad  F_2= -c~\! \mathcal{C}_2,\qquad ~\mathcal{C}_2= \frac{dx_1\wedge dx_2}{x_2^2}+ \text{Vol}(S^2),\\[2mm]
F_4&= \frac{x_2^2}{2} \epsilon_{ijk}\partial_{x_k}(e^{-4A})dx_i\wedge dx_j\wedge\text{Vol}(S^2) ,\qquad \partial^2_{x_1}h+\frac{1}{x_2^2}\partial_{x_2}(x_2^2\partial_{x_2}h)+\partial^2_{x_3}h=0,\nn
\end{align}
where with respect to \eqref{eq: azero PDEssol} we have set $c_1=1$, $c_2=0$, $c_3=c$, the effect of which is a rescaling in $g_s$ and turning off a closed part of the NS 2-form. Notice that the PDE defining these solutions is simply the flat space Laplace equation in five dimensions, with $SO(3)$ rotational symmetry to accommodate the R-symmetry. Now the metric looks formally like the backreaction of a D4 with harmonic function $e^{-4A}$, but there are more fluxes switched on.

\subsubsection{$g=0$ Ansatz}\label{sub:d4d8}
As previously stated, setting $g=0$ puts us in the class of \cite{youm} containing the localised D4-D8 system, albeit with one of the common world volume directions compactified on $S^1$, so this is not new. However, since we will find generalisations of this later, we present the form of the solution here for comparison:
\begin{align}
ds^2_6&= \frac{1}{\sqrt{f}\sqrt{h}}ds^2(S^1) + \frac{\sqrt{h}}{\sqrt{f}}\bigg(dx_1^2+ dx_2^2+ x_2^2  ds^2(S^2)\bigg)+\sqrt{h}\sqrt{f}dx_3^2,\qquad  e^{-4A}= f h,\nn\\[2mm]
F_0&= \partial_{x_3} f,\qquad ~e^{A-\Phi}=f,\qquad ~ f=(c+ F_0 x_3),\nn\\[2mm]
F_4&= \frac{x_2^2}{2} \epsilon_{ijk}dx_i\wedge dx_j \partial_{x_k}h,\qquad \partial^2_{x_3}h+f\bigg(\frac{1}{x_2^2}\partial_{x_2}(x_2^2\partial_{x_2}h)+\partial^2_{x_1}h\bigg)=0,
\end{align}
% subsection 5dsols (end)

\subsection{The $a_1=b=0$ case} % (fold)
\label{sub:4dsols}
In this subsection we will study the $\{b=a_1=0,a_2=1\}$ case.  In this case supersymmetry follows from the following conditions:
\begin{subequations}
\begin{align}
&B_2 = 0,\label{ba10 SUSY a}\\[2mm]
&d(e^{A}w)=0,\qquad d\big[e^{-2A+\Phi}(v_1- e^{C}H_1)\big]=0,\qquad d(e^{2A+C-\Phi})+ e^{2A-\Phi}v_2=0,\label{ba10 SUSY b}\\[2mm]
&d(e^{-\Phi} v_1)\wedge w\wedge \overline{w}=0,\qquad \big[d(e^{2C-\Phi} v_2)-e^{2C-\Phi} H_1\wedge v_1\big]\wedge w\wedge \overline{w}=0\ ,\label{ba10 SUSY c}
\end{align}
\end{subequations}
where $e^{2C} H_1 = d(e^{2C} B_0 )$ (see (\ref{eq:HdB})).
We immediately see from \eqref{ba10 SUSY a} that there is no NS 3-from flux orthogonal to the $S^2$ directions. 
We solve \eqref{ba10 SUSY b} by defining the vielbein and local coordinates
\begin{subequations}
\begin{align}
v_1&= e^{2A-\Phi}\big(dx_1+ \frac{1}{x_2} e^{2C} H_1\big),\qquad v_2=- e^{-2A+\Phi}dx_2,\\[2mm]
 w&= e^{-A}(dx_3+i dx_4),\qquad  x_2 = e^{2A+C-\Phi}\ .
\end{align}
\end{subequations}
We notice that we have $e^{2C}H_1$ appearing in the vielbein for the first time. Since we have fewer equations, we have no freedom to choose which supersymmetry conditions define the vielbein, and thus we cannot avoid this complication as we did in sections \ref{sub:6dsols} and \ref{sub:5dsols}. In what follows we find it useful to decompose the NS form and define the physical fields as
\beq\label{ba10 para}
e^{2C}H_1 = \bigg[\lambda_1 dx_1+\lambda_2dx_2+\lambda_3 dx_3+\lambda_4 dx_4\bigg],\qquad e^{2A-2\Phi}= x_2 h_1,\qquad  e^{-6A+2\Phi}x_2^2=h_2\ .
\eeq
$\lambda_i,h_i$ are functions of $(x_1,\ldots,x_4)$ in terms of which \eqref{ba10 SUSY c}, together with the fact that $ e^{2C} H_1$ should be closed, impose the PDE's
\beq
\partial_{x_i}\lambda_j-\partial_{x_j}\lambda_i=0,\qquad 
\partial_{x_2}((\lambda_1+ x_2) h_1)= \partial_{x_1}(\lambda_2h_1),\qquad 
\partial_{x_1}h_2=x_2 h_1 \lambda_2. \label{ba10 BPD}
\eeq
and no further conditions. In particular there is  a priori no isometry on $M_4$. The metric is
\beq
ds^2_6=e^{4A-2\Phi}\bigg(dx_1+ \frac{1}{x_2} e^{2C} H_1\bigg)^2+  e^{-4A+2\Phi}\bigg(dx_2^2+ x_2^2 ds^2(S^2)\bigg)+ e^{-2A}\bigg(dx_3^2+ dx_4^2\bigg).
\eeq
As before we can extract the  RR fluxes from (\ref{eq:4dflux}) which in terms of \eqref{ba10 para} can be expressed as
\begin{align}\label{eq:4dFlux}
H_3&=e^{2C}H_1\text{Vol}(S^2),\qquad F_0=0,\\[2mm]
F_2&= \big((\lambda_1+x_2)dx_1+ \lambda_2 dx_2\big)\wedge\big(\partial_{x_3}h_1 dx_4-\partial_{x_4}h_1 dx_3\big)-\big(\partial_{x_1}h dx_1+\partial_{x_2}h dx_2\big)\wedge \big(\lambda_3 dx_4-\lambda_4 dx_3\big)\nn\\[2mm]
&+\bigg(\lambda_3 \partial_{x_3}h_1+\lambda_4 \partial_{x_4}h_1-\frac{1}{\lambda_1+x_2}\big(h_1 \partial_{x_1}h_2+ \partial_{x_1}h_1 (\lambda_3^2+\lambda_4^2+ h_2)\big)\bigg)dx_3\wedge dx_4,\nn\\[2mm]
F_4&=\bigg[x_2 h_1dx_1\wedge\big(\lambda_4 dx_3-\lambda_3dx_4\big)+dx_2\wedge\big(\partial_{x_4}h_2dx_3-\partial_{x_3}h_2dx_4\big)\nn\\[2mm]
&+\bigg(x_2^2 h_1\partial_{x_2}(x_2^{-1}h_2)-\frac{1}{\lambda_1+x_2}\big((\partial_{x_1}h_2)^2+ x_2 h_1 (\lambda_3^2+\lambda_4^2+ h_2)\big) \bigg)dx_3\wedge dx_4\bigg]\wedge \text{Vol}(S^2),\nn
\end{align}
where we note that we necessarily have zero Romans mass. Although there are generically fewer RR fluxes turned on here than in the preceding sections, what is present has many legs, so the PDE's following from the Bianchi identities are more involved. Ensuring that $F_2$ is closed imposes three PDEs:
\begin{subequations}\label{eq: F2 bianchi}
\begin{align}
&\partial_{x_1}(\lambda_2 \partial_{x_i}h_1)+ \partial_{x_2}(\lambda_i \partial_{x_1}h_1)=0,\label{eq: F2 bianchi a}\\[2mm]
&\partial_{x_i}\big((\lambda_1+x_2) \partial_{x_i}h_1-\lambda_i\partial_{x_1}h_1\big)= \partial_{x_1}\left(\lambda_i \partial_{x_i}h_1-\frac{1}{\lambda_1+x_2}\big(h_1 \partial_{x_1}h_2+ \partial_{x_1}h_1 (\lambda_i\lambda_i+ h_2)\big)\right),\label{eq: F2 bianchi b}\\[2mm]
&\partial_{x_i}\big(\lambda_2 \partial_{x_i}h_1-\lambda_i\partial_{x_2}h_1\big)= \partial_{x_2}\bigg(\lambda_i \partial_{x_i}h_1-\frac{1}{\lambda_1+x_2}\big(h_1 \partial_{x_1}h_2+ \partial_{x_1}h_1 (\lambda_i \lambda_i+ h_2)\big)\bigg),\label{eq: F2 bianchi c}
\end{align}
\end{subequations}
where $i=3,4$. The $F_4$ Bianchi identity further imposes
\begin{subequations}\label{eq: F4 bianchi}
\begin{align}
&\partial_{x_4}^2h_2+\partial_{x_3}^2h_2+\partial_{x_2}(x_2^2 h_1  \partial_{x_2}(x_2^{-1} h_2))+\partial_{x_2}\bigg[\frac{x_2 h_1}{\lambda_1+x_2}(\lambda_3^2+\lambda_4^2+x_2 \lambda_2^2 h_1+ h_2)\bigg]\nn\\[2mm]
&+ \frac{1}{\lambda_1+x_2}\bigg[\lambda_2 \partial_{x_1}(h_1 h_2)- (\lambda_3^2+\lambda_4^2)((\lambda_1+x_2)\partial_{x_2} h_1- \lambda_2 \partial_{x_1} h_1)\bigg]=0,\label{eq: F4 bianchi a}\\[2mm]
& \partial_{x_4}(x_2 \lambda_4 h_1)+\partial_{x_3}(x_2 \lambda_3 h_1)+ \partial_{x_1}(x_2^2 h_1\partial_{x_2}(x_2^{-1} h_2))- \partial_{x_1}\bigg[\frac{x_2 h_1}{\lambda_1+x_2}(\lambda_3^2+\lambda_4^2+x_2 \lambda_2^2 h_1+ h_2)\bigg]\nn\\[2mm]
&\frac{1}{\lambda_1\!+\!x_2}\bigg[\lambda_1 \partial_{x_1}(h_1 h_2)\!\!-\!\!x_2 (\lambda_3^2\!+\!\!\lambda_4^2) \partial_{x_1}h_1+x_2(\lambda_1+x_2)(\lambda_3 \partial_{x_3} h_1+\lambda_4 \partial_{x_4} h_1)\bigg]=0\label{eq: F4 bianchi b}.
\end{align}
\end{subequations}
Clearly \eqref{eq: F2 bianchi}--(\ref{eq: F4 bianchi}) is a rather complicated system. In fact we can actually get far more compact expressions by performing the coordinate transformation $x_1 \to x_1 -\frac{1}{x_2} e^{2C}B_0$, which we examine in appendix \ref{alt 4dsols}. However these actually turn out to be harder to solve. Despite their complexity we do find some sub classes where the PDEs simplify dramatically, which we will now study.

\subsubsection{$H_1=0$ Ansatz}\label{subsub  H1eq0}
The simplest thing we can do is set all $\lambda_i=0$, so there is no NS flux. This means that \eqref{ba10 BPD} and \eqref{eq: F2 bianchi c} impose
\beq
\partial_{x_2}(x_2 h_1)=\partial_{x_1}h_2= \partial_{x_1}(x_2 h_1)\partial_{x_2}\left(\frac{h_2}{x_2^2}\right)=0\ .
\eeq
This motivates defining
\beq
x_2 h_1= \tilde{h}_1(x_1,x_3,x_4),\quad  h_2=x_2^2 \tilde{h}_2(x_2,x_3,x_4),\quad e^{2A-2\Phi}= \tilde{h}_1,\quad e^{-6A+2\Phi}=\tilde{h}_2,
\eeq
in terms of which the Bianchi identity conditions that are not trivially solved reduce to
\begin{subequations}
\begin{align}
	\partial^2_{x_3}\tilde{h}_1+\partial^2_{x_4}\tilde{h}_1&+ \tilde{h}_2\partial^2_{x_1}\tilde{h}_1=0,\qquad  \partial^2_{x_3}\tilde{h}_2+\partial^2_{x_4}\tilde{h}_2+ \frac{\tilde{h}_1}{x_2^2}\partial_{x_2}(x_2^2 \partial_{x_2}\tilde{h}_2)=0, \label{eq:H10PDE}\\ 
	&(\partial_{x_1}\tilde{h}_1)(\partial_{x_2}\tilde{h}_2)=0.	\label{eq:2op}
\end{align}	
\end{subequations}
The solutions have data
\begin{align}
ds^2_6 &= \sqrt{\frac{\tilde{h}_1}{\tilde{h}_2}}dx_1^2 +\sqrt{\frac{\tilde{h}_2}{\tilde{h}_1}}\left(dx_2^2+x_2^2 ds^2(S^2)\right)+ \sqrt{\tilde{h}_1\tilde{h}_2}\left(dx_3^2+ dx_4^2\right),\nn\\[2mm]
F_2&= (\partial_{x_4}\tilde{h}_1dx_3- \partial_{x_3}\tilde{h}_1dx_4)\wedge dx_1-\tilde{h}_2 \partial_{x_1}\tilde{h}_1 dx_3\wedge dx_4,\qquad H=0, \label{eq:youm}\\[2mm]
F_4&= x_2^2\bigg((\partial_{x_4}\tilde{h}_2dx_3-\partial_{x_3}\tilde{h}_2dx_4)\wedge dx_2+  \tilde{h}_1 \partial_{x_2}\tilde{h}_2dx_3\wedge dx_4\bigg)\wedge \text{Vol}(S^2).\nn
\end{align}
(\ref{eq:2op}) gives two options, which because of (\ref{eq:H10PDE}) require that at least one of $\tilde{h}_1,\tilde{h}_2$ is a flat space harmonic function in $x_3,x_4$ only. If both are, we just reproduce the ``harmonic function rule'' for delocalized branes \cite{Papadopoulos:1996uq,Tseytlin:1996bh,Gauntlett:1996pb}. If  $\tilde{h}_1=\tilde{h}_1(x_3,x_4)$ we have the partially localised solution of D4's ending on D6's presented in \cite[Sec.~4.2]{youm}, specialised to the case where the D6 world volume has an SU(2) isometry. Finally if $\tilde{h}_2=\tilde{h}_2(x_3,x_4)$ the solutions are contained within \cite[Sec.~4.2]{youm} up to performing T-dualities on the three U(1) isometries of  $T^3$.

So in conclusion $H_1=0$ reproduces known intersecting brane solutions only.

\subsubsection{$\lambda_2=0$ Ansatz}
A generalisation of section \ref{subsub  H1eq0} that leads to tractable PDEs is to take only $\lambda_2=0$. Here the supersymmetry conditions \eqref{ba10 BPD} reduces to 
\beq\label{eq:f2zeroSUSY}
\partial_{x_1}( (\lambda_1+x_2) h_1)= \partial_{x_1} h_2=0,\qquad  \partial_{x_2}\lambda_i=0,\qquad  \partial_{x_i} \lambda_j=\partial_{x_j} \lambda_i,
\eeq
while \eqref{eq: F2 bianchi a} requires either $\lambda_3=\lambda_4=0$ or $\partial_{x_1}\partial_{x_2}h_1=0$. We will now consider these two subcases in turn. 

\subsubsection*{The case $\partial_{x_1}\partial_{x_2}h_1=0$:  $\partial_{x_1}$ an Isometry}
Reconciling $\partial_{x_1}\partial_{x_2}h_1=0$ with \eqref{eq:f2zeroSUSY} requires $\partial_{x_1}\lambda_1= \partial_{x_1} h_1=0$. From this it follows that
\beq
h_1 = \frac{\tilde{h}_1(x_3,x_4)}{\lambda_1+x_2},~~h_2=h_2(x_2,x_3,x_4),~~\lambda_i=\lambda_i(x_3,x_4),~~d\lambda_1=0,~~\partial_{x_3}\lambda_4=\partial_{x_4}\lambda_3,
\eeq
where $i=3,4$. We see that $\partial_{x_1}$ is necessarily an isometry. T-dualizing it to IIB produces solutions which are conformal Calabi--Yau type (see footnote \ref{foot:D3}) in a similar way as our discussion in section \ref{sub:a10case}. 

The remaining PDE's \eqref{eq: F2 bianchi}--\eqref{eq: F4 bianchi} then truncate dramatically to
\begin{align}\label{eq: x1isometrysol}
&(\partial_{x_3}^2+\partial_{x_4}^2)\tilde h_1=0,\qquad ~\partial_{x_3}(\tilde h_1^2\lambda_3)+\partial_{x_4}(\tilde h_1^2\lambda_4)=0,\nn\\
&\frac{\lambda_1+x_2}{x_2}(\partial_{x_3}^2+\partial_{x_3}^2)\tilde{h}_2+\frac{h_1}{x_2^2}  \partial_{x_2}(x_2^2\partial_{x_2}\tilde{h}_2)+\frac{2 (\lambda_3^2+\lambda_4^2)\tilde h_1}{x_2(\lambda_1+x_2)^3}=0,
\end{align}
where we have introduced 
\beq
h_2= \frac{x_2}{\lambda_1+x_2}\tilde{h}_2(x_2,x_3,x_4).
\eeq

To make progress, we can now make a separation of variables sub-Ansatz:
\beq
\tilde{h}_2= k(x_2) h_3(x_3,x_4)-l(x_2) h_4(x_3,x_4),\qquad  \frac{(\lambda_1 +x_2)^3 }{x_2}\partial_{x_2}(x_2^2\partial_{x_2}(c_1 k+l))=2\ .
\eeq
We find
\begin{align}
&\lambda_3^2+\lambda_4^2=h_4,\qquad \frac{1}{x_2^2}\partial_{x_2}(x_2^2\partial_{x_2}k)=c_2 \frac{\lambda_1+x_2}{x_2} k,\nn\\[2mm]
&(\partial_{x_3}^2+\partial_{x_4}^2)h_4=0,\qquad (\partial_{x_3}^2+\partial_{x_4}^2)h_3+c_2 h_1(h_3+ c_1 h_4)=0.
\end{align}
We can then parametrise 
\beq
\lambda_3=\cos\alpha \sqrt{h_4},\qquad \lambda_4=\sin\alpha \sqrt{h_4},
\eeq
so that the remaining, non harmonic, PDE's to solve are
\begin{subequations}
\begin{align}
&(\partial_{x_3}^2+\partial_{x_4}^2)h_3+c_2 h_1(h_3+ c_1 h_4)=0 , \label{eq:1pdef20}\\[2mm]
&\partial_{x_4}(\cos\alpha \sqrt{h_4})=\partial_{x_3}(\sin\alpha \sqrt{h_4}),\qquad  \partial_{x_3}(\cos\alpha h_1^2 \sqrt{h_4})=-\partial_{x_4}(\sin\alpha h_1^2 \sqrt{h_4})\ .
\label{eq:2pdef20}
\end{align}	
\end{subequations}
These are hard to make progress with in general, but can be solved. For instance if we change to polar coordinates as $x_3= r\cos\theta$, $x_4=r \sin\theta$, then (\ref{eq:2pdef20}) are solved by
\beq
\alpha= \theta,\qquad  h_4=1,\qquad \tilde h_1= \frac{1}{\sqrt{r}} \cos\left(\frac{\theta}{2}\right),
\eeq
which leaves us with (\ref{eq:1pdef20}) to solve, for example by setting $c_2=0$ so that $h_3$ can be any harmonic function.

\subsubsection*{The case $\lambda_3=\lambda_4=0$:  $\partial_{x_1}$ not an Isometry}
For $\lambda_3=\lambda_4=0$ we define 
\beq
(\lambda_1+ x_2)h_1= \tilde{h}_1(x_1,x_3,x_4),\qquad h_2= (\lambda_1+x_2)^2 \tilde h_2(x_2,x_3,x_4),
\eeq
and then the Bianchi identities impose 
\begin{subequations}
	\begin{align}\label{eq: no isometry}
	&\partial_{x_1} \tilde h_1\partial_{x_2} \tilde h_2=0,\qquad (\partial_{x_3}^2+\partial_{x_4}^2)\tilde h_1+ \tilde h_2\partial_{x_1}^2 \tilde h_1=0,\\
	&(\partial_{x_3}^2+\partial_{x_4}^2)\tilde{h}_2+ \frac{x_2}{(\lambda_1+x_2)^3} \tilde{h}_1\partial_{x_2}\bigg((\lambda_1+x_2)^2 \partial_{x_2}\tilde h_2\bigg)=0.
	\end{align}	
\end{subequations}
If $\partial_{x_1} \tilde h_1=0$, once again $\partial_{x_1}$ is an isometry, which gives a subclass of the solutions we already considered. If on the other hand $\partial_{x_2} \tilde h_2=0$, we are left with the PDEs
\begin{equation}
\begin{split}
	&(\partial_{x_3}^2+\partial_{x_4}^2)\tilde{h}_1+ \tilde{h}_2\partial_{x_1}^2 \tilde h_1=0
	 \ ,\\
	&(\partial_{x_3}^2+\partial_{x_4}^2)\tilde{h}_2=0,
\end{split}	
\end{equation}
These look formally like \cite{youm} again, which we found in (\ref{eq:youm}). We can see, however, that there are more fluxes than in (\ref{eq:youm}); for example, in the present case $H\neq 0$.

%\subsubsection{Enhancement to $SU(2)\times U(1)$ R-symmetry}

% subsection 4dsols (end)

\subsection{The $a_1=0$, $b\neq 0$ case} % (fold)
\label{sub:a10case}
We now study the case where only $a_1=0$, which means that $a_2,b$ are function of the coordinates on $M_4$ such that
\beq \label{eq:ab2}
a_2^2+b^2=1. 
\eeq  
We will assume $b\neq 0$, since $b=0$ was considered in section \ref{sub:4dsols}. We will see on the other hand that the case of section \ref{sub:5dsols} is recovered from the present case as a limit. Finally, we will see that the solutions of this section are in fact related by T-duality to conformal Calabi--Yau-type solutions  \cite{grana-polchinski,giddings-kachru-polchinski,becker2,dasgupta-rajesh-sethi}.

With some care one can establish that the supersymmetry conditions of appendix \ref{app:psp4} follow from
\begin{subequations}\label{a10}
\begin{align}
&d(e^{2A+2C-\Phi} b)+ 2e^{2A+C-\Phi}b v_2=0,\label{a10 a}\\[2mm]
&d\left(\frac{1}{\sqrt{b}}e^{3A/2-\Phi/2}\text{Re}k_1\right)=0,\qquad d\left(\frac{1}{b}e^{-A}(\text{Im}k_2+a_2 B_0v_2)\right)=0,\label{a10 b}\\
&B_2= (a_2\text{Im}k_2+ B_0 v_2)\wedge \text{Re}k_1,\qquad d\left(e^{A-\Phi}\frac{a_2^2}{b}\right)=0,\label{a10 c}\\[2mm]
&d(e^{2A+2C-\Phi}B_0\text{Re}k_2)=0,\qquad d(e^{2A-\Phi}\text{Re}k_2)=0,\label{a10 d}\\[2mm]
&d\left(\frac{1}{b}e^{2A-\Phi}B_0 v_2\wedge \text{Re}k_2\wedge\text{Re}k_1\right)=0,\qquad d\left(\frac{1}{b}e^{A}\text{Re}k_2\right)=0,\label{a10 e}
\end{align} 
\end{subequations}
where the complex 1-forms $k_i$ are the following linear combinations of the vielbein 
\beq\label{ki def}
k_1=b v_1+a w,\qquad k_2= -\overline{a} v_1+b w \ .
\eeq
We use the usual trick of introducing local coordinates to solve \eqref{a10}, this time defining
\begin{align}\label{eq:kvw}
\text{Re}k_1&= \sqrt{b} e^{-3A/2+\Phi/2} dx_1,\qquad  v_2=-e^{-3A-C+\Phi}x_2 dx_2,\qquad \text{Re}k_2= -b e^{-A} dx_3,\\[2mm]
\text{Im}k_2&= -be^{A}(dx_4+ \mathcal{A}),\qquad  x_2 =e^{2A+2C-\Phi} b,\qquad ~\mathcal{A}=-\frac{a_2 }{2 b^{3/2} \sqrt{x_2}}B_0e^{-5A/2+\Phi/2}dx_2,\nn
\end{align}
which implies the vielbein on $M_4$ without loss of generality.
The remaining conditions \eqref{a10 c}--\eqref{a10 e} are then solved uniquely by the surprisingly simple conditions
\beq \label{eq:b2c0}
e^{A-\Phi}= \frac{f(x_3)}{b} ,\qquad  e^{2C} B_0= g(x_3),\qquad \partial_{x_4}A=0,\qquad b^2=\frac{f}{c_0+f},
\eeq
where $c_0$ is an integration constant and we reproduce \eqref{eq:a1a20  BPS} when $c_0=0$. It follows that $\partial_{x_4}$ is an isometry like in section \ref{sub:5dsols}, $a_2,b$ depend on $x_3$ only, and the internal metric is
\beq\label{a10 metric}
ds^2_6=e^{2A} b^2\bigg(dx_4-\frac{a_2 g\sqrt{f}}{b x_2^2} dx_2\bigg)^2+ \frac{e^{-2A}}{f}\bigg( b^2  dx_1^2+  dx_2^2+ fdx_3^2+x_2^2  ds^2(S^2)\bigg).
\eeq
The higher RR fluxes can be derived from \eqref{eq:4dflux} in the same fashion as before and lead to
\begin{align}
B&= g \,\mathcal{C}_2+\frac{a_2 b}{ \sqrt{f}}dx_1\wedge\bigg( dx_4- \frac{a_2 g\sqrt{f}}{b x_2^2}\bigg),\qquad  \mathcal{C}_2= \frac{1}{x_2^2}dx_1\wedge dx_2+ \text{Vol}(S^2)\label{a10 fluxes}\\[2mm]
F_0&=\partial_{x_3}f,\qquad F_2=B F_0- \partial_{x_3}(fg)\mathcal{C}_2,\qquad 
F_4= B\wedge F_2+\frac{1}{2}B\wedge B F_0+\frac{1}{2}\partial_{x_3}(f g^2)\nn\\[2mm]
&+x_2^2\bigg(\frac{1}{b^2}\partial_{x_1}(e^{-4A})dx_2\wedge dx_3-\partial_{x_2}(e^{-4A})dx_1\wedge dx_3+\partial_{x_3}(f^{-1} e^{-4A})dx_1\wedge dx_2\bigg)\wedge\text{Vol}(S^2)\nn\ .
\end{align}

Since $\del_{x_4}$ is an isometry, we can wonder what happens if we T-dualize under it. The structure of the T-dual metric one obtains in IIB suggests now a D3--type solution, since the Mink$_4$ metric is multiplied by $e^{2A}$ and the internal metric has an overall $e^{-2A}$. One can see in fact that the resulting solutions are contained in a famous class  \cite{grana-polchinski,giddings-kachru-polchinski,becker2,dasgupta-rajesh-sethi}. This is most easily seen by looking at the pure spinors. From (\ref{eq:kvw}), (\ref{ki def}) we see $dx_4$ is contained inside $w$, which in turn is in $E_3$ of (\ref{eq:canonical6d vielbein}). So when we T-dualize the pure spinors, $E_3$ in (\ref{eq:pspE}) gets replaced by $e^{1/2 E_3 \wedge \bar E_3}$ and viceversa.\footnote{See \cite[Sec.~6]{gmpt3} and \cite{Grana:2008yw} for more details about T-duality and pure spinors.} We end up with the pure spinors associated to an SU(3) structure. Further analysis reveals they are of the conformal Calabi--Yau type.\footnote{\label{foot:D3} If $\Phi_+ = i e^{iJ}$, $\Phi_-=\Omega$, the IIB version of the pure spinor equations (\ref{eq:psp6d}) results in $d\tilde J=0= d \Omega$, where $\tilde J= e^{2A-\phi}J$; then $\tilde J$ and $\Omega$ define a K\"ahler structure on $M_6$. A solution exists because of Yau's theorem; this is a slight generalization of a Calabi--Yau, which when the dilaton is constant is in fact a conformal Calabi--Yau. The flux $G \equiv f_3 +i e^{-\phi} H$ can be shown to be a primitive $(2,1)$-form; the axiodilaton $\tau=C_0 + i e^{-\phi}$ is holomorphic. See \cite[Sec.~4.3.1]{laces} for more details about how this class derives from the pure spinor equations.} There are however some interesting points about this class, which we will make in section \ref{sub:comp}.

Imposing the Bianchi identities for \eqref{a10 fluxes} leads to
\beq
\partial^2_{x_3}f = 0,\qquad  \partial^2_{x_3}(fg)=0,
\eeq
in common with the conditions of section \ref{sub:5dsols}. The PDE is slightly more general than (\ref{eq: a0 PDE}):
\beq\label{a10 F4 PDE}
\frac{1}{b^2}\partial^2_{x_1}(e^{-4A})+\frac{1}{x_2^2}\partial_{x_2}(x_2^2\partial_{x_2}(e^{-4A}))+\partial^2_{x_3}(f^{-1} e^{-4A})+\frac{1}{x_2^4}\partial^2_{x_3}(f g^2)=0\ .
\eeq
This similarity is also reflected in the fact that \eqref{a10 metric} is a generalisation of \eqref{a1a20 metric}. 

Indeed the $M_6=S^2\times S^1\times M_3$ class of section \ref{sub:5dsols} are a special case of the class in this section; we can simply set $c_0=0$ so that, from (\ref{eq:b2c0}) and (\ref{eq:ab2}), $b=1$ and $a_2=0$. Conversely, given a solution with $c_0=0$, (\ref{eq:b2c0}) gives a way to generate a new family of solutions with $c_0\neq 0$. We will now give two examples of this.

\subsubsection{$\partial_{x_1}^2(e^{-4A})=0$ Ansatz}
If $\partial_{x_1}^2(e^{-4A})=0$ then \eqref{a10 F4 PDE} is trivially independent of $c_0$; but, as we encountered in section \ref{sub:5dsols} for (\ref{eq: a0 PDE}), the PDE is still hard to solve unless either $g=0$ or $F_0=0$. We focus on the former here, as the latter does not require the additional restriction we are now imposing on $e^{-4A}$.
These solutions are of the form
\begin{align}
ds^2_6&=\frac{ b^2}{\sqrt{f h}}dx_4^2+ \frac{\sqrt{ h}}{\sqrt{f}}\bigg( b^2  dx_1^2+  dx_2^2+ fdx_3^2+x_2^2  ds^2(S^2)\bigg),\qquad  b^2=\frac{f}{c_0+f},\nn\\[2mm]
e^{-4A}&=f h,\qquad e^{A-\Phi}= \frac{f}{b},\qquad B=\frac{a_2 b}{ \sqrt{f}}dx_1\wedge dx_4,\qquad F_0=\partial_{x_3} f,\qquad  F_2=F_0 B, \nn\\[2mm]
F_4&=x_2^2\bigg(\frac{f}{b^2}\partial_{x_1}hdx_2\wedge dx_3-f\partial_{x_2}hdx_1\wedge dx_3+\partial_{x_3}hdx_1\wedge dx_2\bigg),\nn\\[2mm]
\partial_{x_1}^2h&=0,\qquad \frac{f}{x_2^2}\partial_{x_2}(x_2^2\partial_{x_2}h)+\partial^2_{x_3}h=0\qquad f=c_1+ F_0 x_3.
\end{align}
We see that for $c_0=0$ we have the behaviour of D8 branes and D4s smeared on $x_1$. By turning on $c_0$ we generate additional $F_2$ and $H_3$ flux. It would be interesting to see if this is somehow related to a known solution-generating technique such as the continuous version of U-duality, but this is currently not clear.

\subsubsection{$F_0=0$ Ansatz}
If $F_0=0$, we can take $f=1$ without loss of generality; $b$ is just a constant, which we can remove from \eqref{a10 F4 PDE} by rescaling $x_1 \to b^{-1}x_1$. This results in solutions of the form
\begin{align}
ds^2_6&= e^{2A}b^2\bigg(dx_4-\frac{a_2 c x_3}{b x_2^2} dx_2\bigg)^2+ e^{-2A}\bigg(dx_1^2+ dx_2^2+ dx_3^2+ x_2^2  ds^2(S^2)\bigg),\nn\\[2mm]
e^{-4A}&=b^4e^{-4\Phi}= \frac{x_2^2 h - c^2}{ x_2^2},\qquad B= c x_3 \mathcal{C}_2+a_2 dx_1\wedge\bigg( dx_4- c\frac{a_2 x_3}{b x_2^2}dx_2\bigg),\qquad  F_2= -c~\! \mathcal{C}_2,\nn\\[2mm]
F_4&= \bigg[\frac{x_2^2}{2 b} \epsilon_{ijk}\partial_{x_k}(e^{-4A})dx_i\wedge dx_j +\frac{a_2 c}{b x_2^2}\big(a_2 c x_3 dx_1\wedge dx_2-b x_2^2 dx_1\wedge dx_4\big)\bigg]\wedge \text{Vol}(S^2),\nn\\[2mm]
&\partial^2_{x_3}h+f\bigg(\frac{1}{x_2^2}\partial_{x_2}(x_2^2\partial_{x_2}h)+\partial^2_{x_1}h\bigg)=0,\qquad \mathcal{C}_2= \frac{1}{b x_2^2}dx_1\wedge dx_2+ \text{Vol}(S^2).
\end{align}
where $h=h(x_1,x_2,x_3)$, and we have fixed $g=c x_3$. For $c_0=0$ and $c\neq0$ we have $F_2$, $F_4$ and $H_3$ fluxes turned on. The effect of turning on $c_0$ is to introduce additional $H_3$ and $F_4$ flux, and to fibre the $x_4$ direction over the rest of the manifold.

% subsection a20case (end)

\subsection{Generic case} % (fold)
\label{sub:generic}
Before moving on to examples, we will study the ``generic case'' where $a_1,a_2,b$ are all non-zero; recall that they should be such that
\beq\label{eq: gen sq cond}
a_1^2+a_2^2+b^2=1\ .
\eeq
The alert reader may ask why we have not discussed the $b=0$ and $a_2=0$ cases separately. In fact, both of these can be obtained as limits of the generic case we study in this section, once we assume that $a_1\neq 0$. (The $a_1= 0$ case was covered previously in section \ref{sub:a10case}.) This is a generalization of the statement we made at the beginning of section \ref{sub:6dsols}, namely that the $b=a_2=0$ case can be obtained as a limit of the generic case treated in this section. 

Inserting the definitions  of \eqref{4d bi} into (\ref{eq: Phip})--(\ref{eq: ImPhim}) for the final time we find the necessary and sufficient conditions for supersymmetry
\begin{subequations}
\begin{align}
&d\left(\frac{a_2}{a_1}e^{-2A}\right) = d\left(\frac{b}{a_1^2}e^{-5A+\Phi}\right)=0,\label{gen SUSY a}\\[2mm]
&d\left(\frac{1}{a_1}e^{-A} k_1\right)=0,\qquad d(e^{2A-\Phi}\text{Re}k_2)=0,\qquad ~ d(e^{4A+C-\Phi}a_1)+ e^{4A-\Phi}a_1 v_2=0,\label{gen SUSY b}\\[2mm]
&d\left(\frac{1}{a_1^2} e^{-4A+\Phi}(\text{Re}k_2- a_1 B_0 v_2)\right)=0,\qquad d(e^{2A+2C-\Phi} (B_0\text{Re}k_2+ a_1 v_2))=0,\label{gen SUSY c},\\[2mm]
&B_2 = \frac{b}{a_1}\big(\text{Im}k_1\wedge\text{Im}k_2-\text{Re}k_1\wedge \text{Re}k_2\big)-\frac{a_2}{a_1}\big(\text{Re}k_1\wedge \text{Im}k_1+\text{Re}k_2\wedge \text{Im}k_2\big)\label{gen SUSY d},
\end{align}
\end{subequations}
where the 1-forms $k_i$ where introduced in \eqref{ki def}.
We notice that \eqref{gen SUSY a} just defines $a_2,b$ in terms of $a_1$ as
\beq\label{eq: a2a1 rule}
a_2= c_0 e^{2A}a_1,\qquad  b=e^{5A-\Phi}\tilde{c}_0 a_1^2 ,
\eeq
where $c_0$, $\tilde{c}_0$ are integration constants. We can solve \eqref{gen SUSY b} without loss of generality by introducing local coordinates $(x_1,\ldots,x_4)$ such that
\begin{align}
v_2&= -\frac{1}{a_1}e^{-4A+\Phi}dx_2,\qquad \text{Re}k_2= e^{-2A+\Phi}dx_1,\nn\\[2mm]
 x_2&= e^{4A+C-\Phi}a_2,\qquad k_1=e^{A}a_1(dx_3+i dx_4),
\end{align}
from which the vielbein on $M_4$ follows using \eqref{ki def}. Then, as \eqref{gen SUSY d} just defines the $S^2$ orthogonal part of the NS 2-form, we are left with only  \eqref{gen SUSY c} to solve. This leads to the PDEs
\begin{subequations}\label{eq: susyPDEsgenral1}
\begin{align}
&\partial_{x_1}\left(\frac{e^{-6A+2\Phi}}{a_1^2}\right)-\partial_{x_2}\left(\frac{e^{-8A+2\Phi}B_0}{a_1^2}\right)=0,\label{eq: susyPDEsgenral1 a}\\[2mm]
&\partial_{x_1}\left(\frac{e^{-8A+2\Phi}B_0 x_2^2}{a_1^2}\right)+ \partial_{x_2}\left(\frac{e^{-10A+2\Phi}x_2^2}{a_1^2}\right)=0,\label{eq: susyPDEsgenral1 b}
\end{align}
\end{subequations}
and informs us that $\partial_{x_3}$ and $\partial_{x_4}$ define two U(1) isometries. Notice that for $a_1=1$ these reproduce the PDEs of \eqref{eq: BPS ba20 a}--\eqref{eq: BPS ba20 b} in section \ref{sub:6dsols}. However now the $T^2$ they form is fibred over the rest, and has a position-dependent modular parameter:
\begin{align}
ds^2_6&=e^{2A}ds^2(\tilde{T}^2)+ \frac{e^{-4A+2\Phi}}{a_1^2}\bigg(\frac{ a_1^2}{a_1^2+b^2}dx_1^2+ e^{-4A}\big(dx_2^2+x_2^2 ds^2(S^2)\big)\bigg),\nn\\[2mm]
ds^2(\tilde{T}^2)&=a_1^2 dx_3^2+(a_1^2+b^2)\bigg(dx_4- \frac{a_2 b e^{-3A+\Phi}}{a_1(a_1^2+b^2)}dx_1\bigg)^2.
\end{align}
We extract the fluxes from  (\ref{eq:4dflux}), which, after some significant massaging, can be expressed as
\begin{align}
&B=\frac{e^{-8A+2\Phi}}{a_1^2} x_2^2 B_0 \text{Vol}(S^2)- e^{2A} a_1 dx_3\wedge \bigg(a_2 dx_4 + \frac{b}{a_1} e^{-3A+\Phi}dx_1\bigg),~~F_0=2 e^{-2\Phi}\frac{1}{a_1^2}\partial_{x_1}(e^{2A}a_1),\nn\\[2mm]
&F_2=B F_0 +d\bigg(e^{A-\Phi} b\bigg)\wedge dx_3- x_2^2 \bigg(\partial_{x_2}\left(\frac{a_1^2+b^2}{a_1^2} e^{-4A}\right)-e^{-2A}B_0\partial_{x_1}\left(\frac{a_1^2+b^2}{a_1^2} e^{-4A}\right)\bigg),\nn\\[2mm]
&F_4=B\wedge F_2-\frac{1}{2}B\wedge B F_0-d\bigg(e^{-7A+\Phi}x_2^2 B_0 \frac{b}{a_1^2}\bigg)\wedge dx_3\wedge \text{Vol}(S^2) \ .
\end{align}
Although the fluxes are rather complicated they actually only give rise to a single PDE when we impose their Bianchi identities, namely
\beq\label{eq: gen bianchi}
 \partial_{x_2}\left(\frac{a_1^2+b^2}{a_1^2} e^{-4A}\right)-e^{-2A}B_0\partial_{x_1}\left(\frac{a_1^2+b^2}{a_1^2} e^{-4A}\right)=-\frac{c_1}{x_2^2},
\eeq
for some constant $c_1$. So we are left with three PDEs to solve: \eqref{eq: susyPDEsgenral1} and \eqref{eq: gen bianchi}, much like in section \ref{sub:6dsols} we had (\ref{eq: BPS ba20}) and (\ref{ba20 bianchi}). It will once more turn out that stipulating whether $F_0=0$ or not will reduce these three to one. The similarity actually goes much further: we can show that any solution of section \ref{sub:6dsols} implies a solution of the generic class.

\subsubsection{$F_0=0$}
For $F_0=0$  we need $a_1 e^{2A}$ to be a function of $x_2$ only which implies through \eqref{eq: gen sq cond} that the same is true of  $a_2$ and $a_1^2+b^2$. The Bianchi identity \eqref{eq: gen bianchi} is then solved by
\beq
e^{-4A}= \left(\frac{a_1^2}{a_1^2+b^2}\right)e^{-4A_0},\qquad  e^{-4A_0}= c_2+ \frac{c_1}{x_2}.
\eeq
We can then take \eqref{eq: susyPDEsgenral1 a} as an integrability condition implying the existence of a function $h(x_1,x_2)$ such that
\beq
a_1e^{2\Phi}= (a_1^2+ b^2)^{3/2} e^{6A_0}\partial_{x_1}h,\qquad a_1 B_0=e^{2A_0}\sqrt{a_1^2+ b^2}\frac{\partial_{x_2}h}{\partial_{x_1}h}.
\eeq
Equipped with these definitions \eqref{eq: a2a1 rule} becomes
\beq
a_1^2+b^2= \frac{1}{1+c_0^2 e^{4A_0}},\qquad a_2= \frac{c_0 e^{2A_0}}{\sqrt{1+c_0^2 e^{4A_0}}},\qquad  b= \frac{\tilde{c}_0 e^{2A_0}}{\sqrt{\partial_{x_1} h}\sqrt{1+c_0^2 e^{4A_0}}},
\eeq
which satisfy \eqref{eq: gen sq cond}, while \eqref{eq: susyPDEsgenral1 a} reduces to
\beq \label{eq:gen}
\frac{1}{x_2^2}\partial_{x_2}(x_2^2 \partial_{x_2} h)+ e^{-4A_0}\partial^2_{x_1}h=0, 
\eeq
which is identical to \eqref{eq: h7dcond} with $A\to A_0$. So we see that any massless solution of section \ref{sub:5dsols} implies a massless solution with $c_0$ and/or $\tilde{c}_0$ turned on.

\subsubsection{$F_0\neq 0$}

When $F_0\neq0$ we can define the dilaton in terms of the Romans mass and take  \eqref{eq: gen bianchi} as a definition of $B_0$ so that
\beq
e^{2\Phi}=\frac{2}{a_1^2 F_0 } \partial_{x_1}(a_1e^{2A}),\qquad B_0 =e^{2A}\frac{ c_1+ x_2^2 \partial_{x_2}\left(\frac{1}{a_1^2}e^{-4A}\right)}{x_2^2\partial_{x_1}(\frac{1}{a_1^2}e^{-8A})}.
\eeq 
We use these to substitute for $B_0$ and $\Phi$ in favour of $A$  in \eqref{eq: susyPDEsgenral1 a}--\eqref{eq: susyPDEsgenral1 b} which reduce to a single PDE
\beq\label{eq: PDEA2}
\frac{1}{x_2^{2}}\partial_{x_2}\bigg(x_2^2\partial_{x_2}\left(\frac{a_1^2}{a_1^2+ b^2}e^{-4A}\right)\bigg)+\frac{1}{2}\partial_{x_1}\bigg(\partial_{x_1}\left(\frac{a_1^4}{(a_1^2+ b^2)^2}e^{-8A}\right)\bigg)=0.
\eeq
Just like the $F_0=0$ case we can then define
\beq 
e^{-4A}= \frac{a_1^2}{a_1^2+ b^2} e^{-4A_0},
\eeq
where we now have $A_0(x_1,x_2)$ and see that \eqref{eq: PDEA2} just reduces to \eqref{eq: PDEA} with $A\to A_0$. In other words one can take any massive solution of section \ref{sub:5dsols} and it will imply a massive solution in this more general context.

% subsection generic (end)

\subsection{Summary of this section} % (fold)
\label{sub:sum}

\begin{figure}[ht]
	\centering
		\includegraphics[height=3in]{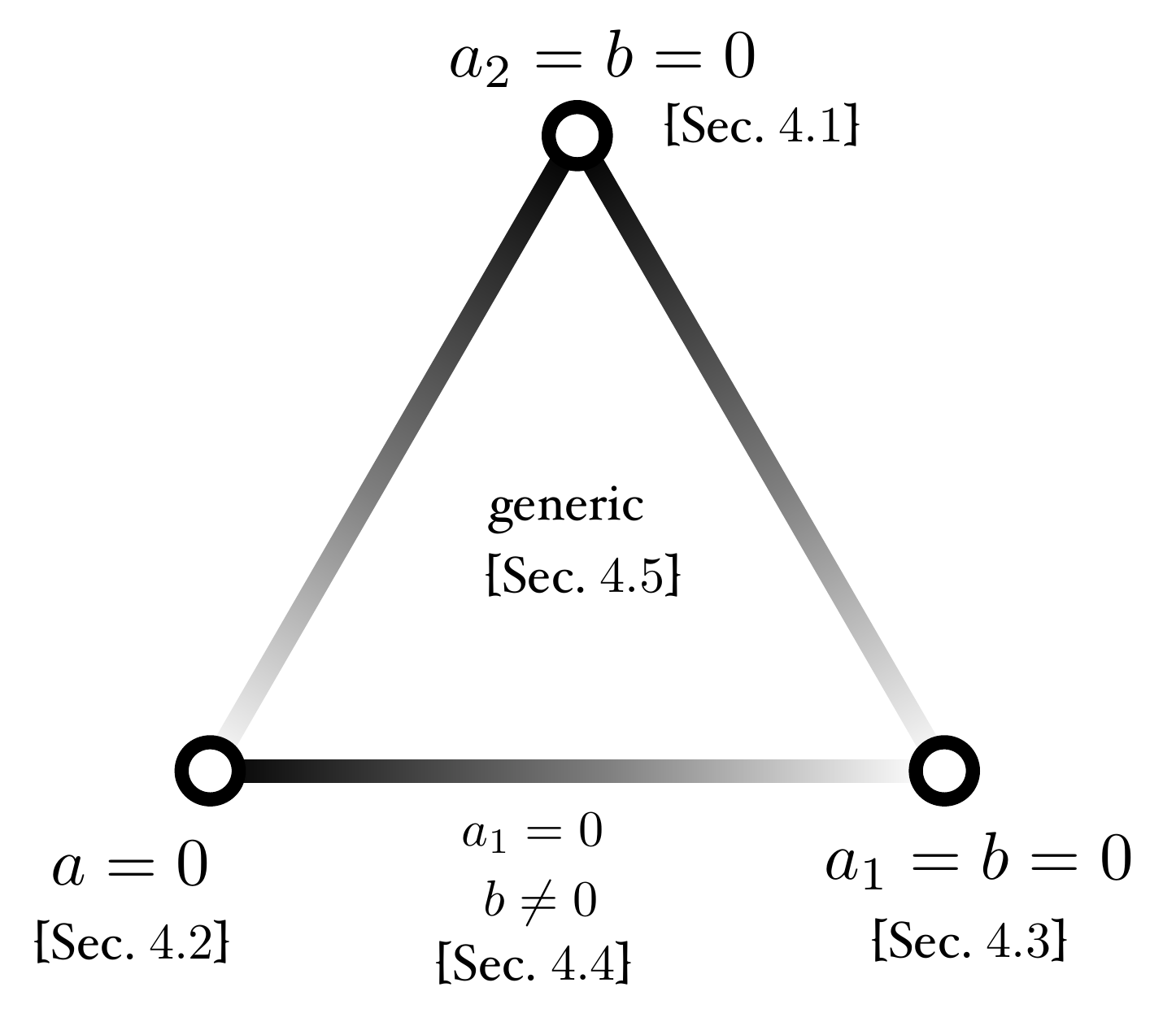}
	\caption{A summary of the classification in this section. The interior of the triangle represents the generic case where none of the parameters vanish, and the sides and vertices represent the various particular cases; recall $a_1\equiv {\rm Re} a$, $a_2 \equiv {\rm Im} a$. The shading of the sides is meant to suggest which limits reproduce the cases represented by the vertices. For example, if one takes the equations for the $a_1=0$ case (lower side) and takes the $a_2\to 0$ limit, one recovers the equations of the $a=0$ case, while the same is not true if one takes the $b\to 0$ limit.}
	\label{fig:sum}
\end{figure}

We summarize the classification we obtained in this section in figure \ref{fig:sum}. Recall $a_1\equiv {\rm Re} a$, $a_2 \equiv {\rm Im} a$.

Strictly speaking one only needs to distinguish three cases. 
\begin{itemize}
	\item Generic case: $a_1\neq 0$ (section \ref{sub:generic}). In this case one needs to solve the PDEs \eqref{eq: susyPDEsgenral1} and \eqref{eq: gen bianchi}. In the case with and without $F_0$, these reduce respectively to either (\ref{eq:gen}) or (\ref{eq: PDEA2}). The solutions can be viewed as a certain decoration by extra fluxes of the particular case $a_2=b=0$ (upper vertex in figure \ref{fig:sum}, section \ref{sub:6dsols}), where $M_6=T^2\times S^2\times M_2$ and which can \emph{roughly} be interpreted as systems of NS5s, D6s and D8s. 
	\item $a_1=0$, $b\neq 0$ (section \ref{sub:a10case}; lower side in figure \ref{fig:sum}). Here one needs to solve (\ref{a10 F4 PDE}). There is in this class always an abelian isometry, T-dualizing along which produces a conformal Calabi--Yau-type solution \cite{grana-polchinski,giddings-kachru-polchinski,becker2,dasgupta-rajesh-sethi} (conformal Calabi--Yau's with $(2,1)$ primitive flux). Again one can view the solutions as a decoration by extra fluxes of the particular case $a=0$ (lower-left vertex in figure \ref{fig:sum}, section \ref{sub:5dsols}), where $M_6=S^1\times S^2\times M_3$.
	\item $a_1=b=0$ (section \ref{sub:4dsols}). In this case the system is the rather more complicated \eqref{eq: F2 bianchi}--(\ref{eq: F4 bianchi}), although we did manage to simplify it considerably and solve it with several Ans\"atze. Among these, we found some known D4--D6 systems, and some generalizations thereof. This case is in a sense the most interesting one.
\end{itemize}

We will see now some examples.

% subsection sum (end)

% section class (end)

\section{Examples} % (fold)
\label{sec:ex}

In section \ref{sec:class} we classified solutions to our minimal class, as defined in section \ref{sec:ansatz}. The classification was organized in several cases; in each of these the requirement of preserved supersymmetry was reduced to a system of PDEs. In this section, we will look at some particular solutions of these systems. In section \ref{sub:comp} we will look at cases where $M_6$ is compact, namely at Minkowski compactifications. In later subsections we will do the exercise of applying the systems to AdS solutions. In particular, we will recover the AdS$_7$ and AdS$_5$ solutions of \cite{afrt,10letter} and \cite{lin-lunin-maldacena}, and we will write AdS$_6$ solutions in terms of a Toda-like equation which presumably reproduces the solutions in \cite{dhoker-gutperle-uhlemann,DHoker:2016ujz}. For AdS$_5$, the only solutions in our class will actually be those of \cite{lin-lunin-maldacena}.

\subsection{Compact $M_6$} % (fold)
\label{sub:comp}

We will now try to find examples where $M_6$ is compact, in some of the classes considered in \ref{sec:class}. The classification there produced a vast array of possible systems, and our analysis here should be regarded as preliminary. 

\paragraph{[$b=a_2=0$.]} Let us start from the $b=a_2=0$ case of section \ref{sub:6dsols}. Specifically, we will look at the $F_0=0$ case, as in section \ref{sub:ima0}. We reduced there the problem to the single PDE (\ref{eq: h7dcond}). We collect the data of the solution here:
\begin{equation}\label{eq:ima1}
\begin{split}
	ds^2&= e^{2A} ds^2_{{\rm Mink}_4\times T^2}+ \del_{x_1} h \left( e^{2A} dx_1^2+ e^{-2A}( dx_2^2+x_2^2 ds^2(S^2))\right)\ ,\\
	e^{2\Phi}&=e^{6A} \del_{x_1} h\ ,\qquad B= x_2^2\del_{x_2} h \text{Vol}(S^2)\ ,\qquad F_2= c_1 \text{Vol}(S^2)\ .	
\end{split}	
\end{equation}
To find solutions, we can impose a separation of variables Ansatz $h= f(x_1) g(x_2)$. Then \eqref{eq: h7dcond} imposes
\beq\label{eq: massless sep ansatz}
\partial_{x_1}^2f+ c f=0 \ ,\qquad \partial^2_{x_2}( x_2g)= c(c_1+c_2 x_2)g,
\eeq
which have closed form solutions. For $c=0$ we have compact solutions for $h=C x_1 \frac{a-x_2}{x_2}$, $e^{4A}=x_2/c_1$, if we take $x_1$ to be periodically identified with itself. So (\ref{eq:ima1}) becomes 
\begin{equation}\label{eq:D6smNS5}
\begin{split}
	ds^2&= \sqrt{\frac{x_2}{c_1}} ds^2_{{\rm Mink}_4\times T^2}+ C\left(\frac a{x_2}-1\right)\left(\sqrt{\frac{x_2}{c_1}} dx_1^2+ \sqrt{\frac{c_1}{x_2}}( dx_2^2+x_2^2 ds^2(S^2))\right)\ ,\\
	e^{2\Phi}&=\frac{C}{c_1^{3/2}} \sqrt{x_2}(a-x_2)\ ,\qquad H= a C dx_1\wedge \text{Vol}(S^2)\ ,\qquad F_2= c_1 \text{Vol}(S^2)\ .	
\end{split}	
\end{equation}
If the factor $\big(\frac a{x_2}-1\big)$ were $\left(1+\frac a{x_2}\right)$, this would represent a near-horizon D6 with an NS5 smeared along the $x_1$ direction. With $\left(1-\frac a{x_2}\right)$, the NS5 would be replaced by a more peculiar object, an ONS5-plane.\footnote{This is an NS analogue of an O-plane, related at the worldsheet level \cite{Sen:1996na,Sen:1998ii} to an inversion of four coordinates times the left fermionic number $(-)^{F_L}$ (rather than $\Omega$ as for usual O-planes). It can also be thought of as the S-dual of an O5.} As it is, (\ref{eq:D6smNS5}) is a compact solution which behaves close to $x_2=0$ as a D6/NS5, and close to $x_2=a$ as a D6/ONS5, both NS objects being smeared along $x_1$. The solution is however highly curved everywhere: playing with the parameters (and taking flux quantization into account) one cannot make $a$ large. Solutions of this type, obtained by reversing signs in otherwise well-known solutions with orientifold planes, are easy to obtain, and we could exhibit several other such examples for other classes in section \ref{sec:class}. We will not comment on such examples further. Notice however that the present example becomes much less trivial after one applies to it the solution-generating technique of section \ref{sub:generic}.

For $c>0$, (\ref{eq: massless sep ansatz}) gives more interesting results. It can again be solved explicitly: 
\begin{equation}
\begin{split}
		f&= -\cos(\sqrt{c} x_1) \ ,\\ g&= e^{-\sqrt{c c_2} x_2}\left(\alpha {}_1F_1\left(1+\frac{c_1}2 \sqrt\frac c{c_2},2,2\sqrt{c c_2}x_2\right) + \beta U\left(1+\frac{c_1}2 \sqrt\frac c{c_2},2,2\sqrt{c c_2}x_2\right) \right)
\end{split}
\end{equation}
with $\alpha$ and $\beta$ two real parameters. For example, let us assume $c_2>1$, $c_1<0$, so that $e^{-4A}=c_2 + \frac{c_1}{x_2}$ from (\ref{eq:e-4Aima}) corresponds to an O6 at $x_2=x_2^-=-\frac{c_1}{c_2}$. Now $\del_{x_1} h$ in (\ref{eq:ima1}) is $\sqrt{c} \sin(\sqrt{c} x_1) g(x_2)$. The fact that $\sin(\sqrt{c} x_1)$ has a simple zero at $x_1=0,\frac\pi{\sqrt{c}}$ signals the presence at those loci of NS5s smeared along $x_2$ and the $S^2$. Moreover, by adjusting the $\beta/\alpha$, one can arrange for $g$ to have a simple zero at some finite value $x_2=x_2^+$. This signals the presence there of an ONS5. This time we then take $x_1 \in [0,\frac\pi{\sqrt c}]$, $x_2 \in [x_2^-,x_2^+]$. Flux quantization fixes $c_1$ and $\alpha$. It appears to be possible to achieve weak curvature and weak coupling by adjusting the remaining parameters. The presence of so many sources, one of which a bit exotic, should make one cautious. But this is only one possible example of this type; clearly it is promising to explore such solutions further.

\paragraph{[$a=0$.]} As we have stressed earlier, this subclass is T-dual to the well-known conformal Calabi--Yau-type solutions in \cite{grana-polchinski,giddings-kachru-polchinski,becker2,dasgupta-rajesh-sethi} (conformal Calabi--Yau's with $(2,1)$ primitive flux). 

However, even in this class we find something interesting. Let us consider (\ref{eq:a0F0}): as we observed there, the metric is formally of the D4-brane type. If we choose the particularly simple harmonic function $h=4(a-x_2)$ and the constant $c=a \sin\theta_0$, with $a>0$, $\theta_0\in (0,\pi/2)$, the Minkowski warp factor in (\ref{eq:a0F0}) is 
\begin{equation}
	e^{-4A}=\frac{(a_+-x_2)(x_2-a_-)}{x_2^2} \ ,\qquad a_{\pm}=\frac{a}{2}(1\pm\cos\theta_0) \ .
\end{equation}
The coordinate $x_2$ can be taken to belong in the range $(a_-,a_+)$; at both extrema, the metric's behavior is that of an O4 smeared along directions $x_1$ and $x_3$. The fluxes read
\begin{equation}
\begin{split}
	\label{eq: simple R15 2}
	H&= a L^2\sin\theta_0 dx_3\wedge{\cal C}_2 \ ,\qquad F_2=-\frac{a L^2}{c_1^{3/2}} \sin\theta_0 {\cal C}_2, \\ F_4&= \frac{4a L^4}{c_1^{3/2}}\left(1-\frac{a}{2 x_2}\sin^2\theta_0\right)dx_1\wedge dx_3\wedge\text{Vol}(S^2)\ .	
\end{split}	
\end{equation}
where ${\cal C}_2$ is defined in \eqref{eq:a0F0}. As remarked earlier, if we T-dualize this example along $x_4$ we obtain a D3-type solution; technically, the internal space is a conformal Calabi--Yau. However, this solution should not be interpreted, as is usually done for such solutions, as the result of backreaction of O3-planes and D3-branes on a preexisting Calabi--Yau. The internal space $M_6$ looks rather like a $T^3\times S^3$ with two O3-planes at the two poles of the $S^3$, smeared along the $T^3$. The reason this falls in the conformal Calabi--Yau class is the accident that $S^3$ is itself conformally flat. If one is (rightly) not happy about the fact that the O3-planes are smeared, one can T-dualize instead along the directions $x_1$, $x_3$, to obtain a IIA solution again with $M_6=T^3\times S^3$, but with O6-planes extended along the $T^3$ and with two O6-planes  at the poles of the $S^3$. Moreover, we have studied flux quantization and we have found no obstruction to making the solution weakly curved and weakly coupled (away from the two O6-planes), although we refrain from giving the results here.

% subsection comp (end)

\subsection{AdS$_7$} % (fold)
\label{sub:ads7}

We will now show how to recover the AdS$_7$ solutions of \cite{afrt,10letter} as particular examples of the system in section \ref{sub:imamass}. As we explained there (see footnote \ref{foot:ima}), the relevant equation (\ref{eq: PDEA}) is in fact the one derived in \cite{imamura-D8,janssen-meessen-ortin}. A relation between those papers and the AdS$_7$ solutions was conjectured already in \cite{afrt}, but never realized until now.\footnote{We remind the reader that while this work was being completed, \cite{bobev-dibitetto-gautason-trujien} appeared, which has some overlap with the results of this section.}

We will use the customary trick of  viewing AdS$_7$ solutions as Mink$_6$ solutions (mentioned in (\ref{eq:mink-ads}) for AdS$_5$). Thus in (\ref{ba20 6dmet}) we replace $T^2$ by $\R^2$, and we take $e^A= e^{\rho+A_7}$. We then try to rewrite the metric for the remaining four directions as 
\begin{equation}\label{eq:43f}
	ds^2_4= e^{2A_7} d \rho^2 + ds^2_3\ .
\end{equation}
In order to do so, we define the following change of variables: 
\beq \label{eq:x12fg}
x_1 = f(z) e^{2\rho},\qquad x_2 = g(z) e^{4\rho}\ .
\eeq
Imposing that (\ref{ba20 6dmet}) does indeed look like (\ref{eq:43f}) gives two conditions: one from the $d\rho^2$ coefficient, one from setting to zero the coefficient of $d\rho dz$. This results in   
\beq\label{eq: AdS7cond}
\partial_{z}f= \frac{16 g^2 \partial_{z}(e^{-4A_7})}{8f- e^{4A_7} F_0}\ ,\qquad\partial_{z}g=- e^{4A_7}\frac{8 f g \partial_{z}(e^{-4A_7})}{8f- e^{4A_7} F_0}\ .
\eeq
With these definitions, it turns out that eq \eqref{eq: PDEA} is solved automatically. We can then use reparametrisation invariance to choose
\beq
g=  e^{4A_7}\sqrt{ \frac{1}{8}( \frac{3 c_1 F_0^2}{16}- F_0 f)}\ ,
\eeq
which reduces \eqref{eq: AdS7cond} to a single ODE:
\beq
\partial_{z} (e^{4A_7})= \frac{8 \partial_{z}f(F_0e^{4A_7}-8 f )}{F_0(3 c_1 F_0 -16 f)}\ .
\eeq
This is solved by
\beq
e^{4A_7}= c_1+\frac{8 f}{3 F_0}+ \frac{4 \tilde{c}_2}{F_0\sqrt{16f-3 c_1 F_0}}\ .
\eeq
where $\tilde{c}_2$ is an integration constant.
Since $e^{4A_7}$ is actually just a function of $f$, we can choose $f$ such that expressions simplify. Taking 
\beq
f= \frac{3 F_0}{8} \left(\frac{c_1}{2}+ z^2\right),
\eeq
and redefining $\tilde{c}_2$ in terms of a new constant $c_2$ we get
\begin{equation}
	 e^{4A_7}= \frac{2z^3 + 3c_1 z + 2 c_2}{2z}\ ,
\end{equation}
which is proportional to $\frac{\alpha}{\ddot \alpha}$, for $\alpha$ a cubic function. This is the expression for $e^{4A_7}$ one can find in \cite[Eq.~(2.27)]{cremonesi-t}; $\alpha=\sqrt \beta$ can be found in (2.26) there and is indeed a cubic function. So our $z$ here is linearly related to the $z$ there. One can also check that the local expression for the dilaton, metric and fluxes that we obtain from section \ref{sub:imamass} reproduce those in \cite[Eq.~(2.27)--(2.29)]{cremonesi-t}. 

% subsection ads7 (end)

\subsection{AdS$_6$} % (fold)
\label{sub:ads6}

We will now use a similar logic to the problem of AdS$_6$ solutions in IIB using section \ref{sub:4dsols}.\footnote{There is also a solution in IIA \cite{brandhuber-oz}, found to be unique in \cite{passias}. This can be easily reproduced as a particular case of the system in section \ref{sub:d4d8}; we will not present the details here.} These have been treated in \cite{afprt}, where the problem was reduced to two hard PDEs for the warping and dilaton (which were later confirmed in \cite{Kim:2015hya}). Recently, the general solution was found in \cite{dhoker-gutperle-uhlemann,DHoker:2016ujz} by reanalyzing the problem in a clever coordinate system.  

Since we are actually interested in IIB solutions, our Ansatz will correspond to a formal T-dual of such a solution along one of the spacetime directions:
\beq \label{eq:Tads6}
ds^2= 4e^{2A_6}\bigg(e^{2\rho} dx^2_{1,3}+ d\rho^2\bigg)+ \frac{1}{4}e^{-2\rho-2A_6}dx_4^2 + ds^2(M_4),~~~e^{\Phi}= e^{\Phi_{\rm IIB}-\rho-A_6}  \ ;
\eeq
Thus in particular $e^{2A}= 4e^{2A_6+2\rho}$, where the 4 is inserted for convenience as it simplifies later expressions. $M_4$ is spanned by $S^2$ and two other coordinates. $x_4$ is the direction along which T-duality will produce an AdS$_6$ solution in IIB. The fluxes can be found in \eqref{eq:4dFlux} from which one sees that while $\partial_{x_4}$ is not an isometry in general, it is possible to impose that it is. We will try to rewrite the metric of section \ref{sub:4dsols} in this form so that it reduces to (\ref{eq:Tads6}). Similarly to (\ref{eq:x12fg}), we will change coordinates:
\beq
x_1 = e^{-3\rho} f(r,y)\ ,\qquad x_2 = \frac{4}{9} e^{3\rho} y \ ,\qquad x_3 =4 r e^{\rho-\frac{1}{3}\Delta(r,y)}\ .
\eeq
We correspondingly write $e^{2C}H_1=d(e^{2C} B_0)$ (recall (\ref{eq:Bdec})) as
\beq
e^{2C}H_1= h_1(r,y)dr+ h_2(r,y)dy\ ,
\eeq
Imposing that the metric has no $d\rho$ cross terms and that the coefficient of $d \rho^2$ is correct leads to 
\begin{equation}
\begin{split}
e^{\Phi_{\rm IIB}}&= 6 \frac{e^{-\frac{2}{3}\Delta+4 \lambda}+ e^{12 \lambda} f^2}{e^{8\lambda}- y^2},\\
	h_1&=\frac{4 y}{9f}\bigg(r(3-r \partial_r \Delta)e^{-\frac{2}{3}\Delta-8 \lambda}- f\partial_r f \bigg) \ ,\qquad h_2=\frac{4 y}{9} ( f \partial_y \Delta- \partial_y f)\ ,
\end{split}	
\end{equation}
where to simplify things we have made the choice
\begin{equation}
	e^{4A_6}= \frac{1}{6}e^{4\lambda+\Phi_{\rm IIB}}\ .
\end{equation}
Diffeomorphism invariance now gives us the freedom to choose $\Delta$ such that the metric also has no $drdy$ cross term,  which actually defines the warp factor in terms of $\Delta$ as
\beq
e^{-8\lambda}=\frac{\partial_{y}\Delta}{y(1+ y\partial_{y}\Delta)}.
\eeq
This leaves us with two functions, $f$ and $\Delta$ which the physical fields are defined in terms of, but we are yet to impose the supersymmetry constraints (\ref{ba10 SUSY c}), these lead to
\beq
f=-e^{\frac{1}{3}\Delta}\frac{3-r \partial_r\Delta}{(1+ y\partial_y \Delta)} \ ,\qquad\partial^2_r (e^{\frac{1}{3}\Delta})= \frac{1}{3}\partial_{y}^2 e^{-\Delta}\ ,
\eeq
from which it follows that the NS 3-form is closed and that the Bianchi identities of the RR sector are all satisfied. The second condition is reminiscent of (but not the same as) a Toda equation, such as the one we will review in section \ref{sub:ads5}.\footnote{\label{foot:strictly} Strictly speaking, the supersymmetry conditions impose that
$\partial_y \log(f)= \partial_{y} \log\left(-e^{\frac{1}{3}\Delta}\frac{3-r \partial_r\Delta}{(1+ y\partial_y \Delta)}\right)$
which means we can multiply our definition of $f$ by any function $k(r)$ and still satisfy supersymmetry and the Bianchi identities. This is a manifestation of diffeomorphism invariance, so we can set $k$ to whatever we choose and still describe the same physical system. We encounter the same situation when we impose supersymmetry for the $AdS_5$ solutions in section \ref{sub:ads5}: there only one choice of this arbitrary function leads to the well known Toda result.} It is much simpler than the PDEs in \cite{afprt} ; it should be solved by the metrics in \cite{dhoker-gutperle-uhlemann,DHoker:2016ujz}. T-dualizing to IIB, the metric reads 
\beq
ds^2= 4 e^{2A_6}\bigg(ds^2_{\rm AdS_6}+\frac{y \partial_y\Delta}{9(1+y \partial_y\Delta)}ds^2(S^2)+ \frac{\partial_{y}\Delta}{9y}\big(dy^2+ e^{-\frac{4}{3}\Delta} dr^2\big)\bigg)\,
\eeq
while the fluxes on the other hand
\begin{align}
F_1&= \bigg(\frac{27}{4y^2} e^{-2/3\Delta+8\lambda-2\Phi}f h_2- \frac{2}{3} e^{-2/3\Delta-\lambda-7/4\Phi}\partial_y(y e^{-3\lambda+ 3/4 \Phi}) \bigg)dr\nn\\[2mm]
&-\bigg(\frac{27}{4 y^2}e^{2/3\Delta+8\lambda-2\Phi}f h_2- \frac{2}{3}e^{2/3\Delta-\lambda-7/4\Phi}\partial_r(y e^{-3\lambda+ 3/4 \Phi})\bigg) dy,\\[2mm]
F_3&=\frac{y}{54}e^{-2/3\Delta-4\lambda-\Phi}\bigg((45 h_2 +32 y \partial_y f+ 96 y f \partial_y \lambda) dr+(45 h_1 +32 y \partial_r f+ 96 y f \partial_r \lambda)e^{4/3\Delta}dy\bigg)\wedge \text{Vol}(S^2)\nn.
\end{align}
A particularly simple solution is given by
\beq
e^{\Delta}=  \frac{c_1 r^3}{c_2- y},
\eeq
which gives rise to the Hopf fibre T-dual of the unique $AdS_6$ solution in IIA. As there is no uniqueness theorem in IIB, and indeed examples beyond the IIA T-dual are already known, our hope is that this formulation will prove useful to further populate the class of such solutions. We leave this however for future work.

\subsection{AdS$_5$} % (fold)
\label{sub:ads5}

We will now also show how to recover in IIA the AdS$_5$ solutions obtained in M-theory by \cite{lin-lunin-maldacena} from section \ref{sub:4dsols}.  

The strategy is similar as the one we used in the previous subsections, and we will be brief. Once again we use the trick (\ref{eq:mink-ads}); however, we actually define $e^{2A} = 4 e^{2\rho+ 2A_5}$, to match the conventions in \cite{lin-lunin-maldacena}. These solutions have an enhanced SU(2)$\times$ U(1) R-symmetry which should be manifest in the geometry in appropriate coordinates. If we impose rotational symmetry in $(x_3,x_4)$ we introduce a U(1) isometry in the metric under which the pure spinors are charged, so this is the U(1) R-symmetry. This motivates the change of coordinates
\beq
x_1= f(r,y)e^{-2\rho} \ ,\qquad  x_2= 4y e^{2\rho} \ ,\qquad  x_3+ ix_4 = 4r e^{\frac{1}{2}D(r,y)} e^{i\psi} e^{\rho},
\eeq
where $\partial_{\psi}$ is part of the R-symmetry and the factors of $\rho$ are chosen such that the internal manifold has no $\rho$-dependent warping. We then express 
\beq
e^{2C}H_1=  h_1(r,y)dr+ h_2(r,y) dy,
\eeq
which is the most general form consistent with the isometries. To obtain $AdS_5$ solutions, we also need to again impose that the cross-terms involving $d \rho$ vanish, and that the coefficient of $d \rho^2$ is correct. By imposing these conditions and that the metric is diagonal in all coordinates we fix the physical fields as
\begin{align}
e^{2A_5}&= e^{2\lambda+ 2/3\Phi} \ ,\qquad e^{-6\lambda}=\frac{-\partial_{y} D}{y(1-y \partial_yD)} \ ,\qquad e^{4/3\Phi-2\lambda}=\frac{r^2 e^{D}+16 f^2e^{6\lambda} }{e^{6\lambda}-y^2} ,\\[2mm]
h_1&=r y e^{D-6\lambda}\frac{2+r \partial_rD}{4f}-4y \partial_r f \ ,\qquad 
h_2= \frac{y}{4 f} \left(e^{-8 \lambda+4/3\Phi}y+ e^{D-6\lambda}r^2\partial_y D-16 f \partial_yf\right)\ ,
\end{align}
where the first of these just defines $A_5$ in terms of an arbitrary function $\lambda(r,y)$. We now have two undefined functions $f$ and $D$, but we are yet to impose the conditions (\ref{ba10 SUSY c}); these lead to
\beq
f=\frac{2+r \partial_rD}{4(1- y\partial_y D)}\ ,\qquad \frac{1}{r}\partial_r(r\partial_rD)+ \partial_{y}^2 e^{D}=0\ .
\eeq
up to the subtlety discussed in footnote \ref{foot:strictly}.
Thus, $D$ must obey the axially symmetric 3d Toda equation, and $f$ is fixed in terms of $D$. With these conditions the NS three-form is closed, and the fluxes obey the Bianchi identities.
The solutions are then of the form
\begin{align}\label{eq:Ads5sol}
ds^2&= e^{2\lambda+ \frac{2}{3} \Phi}\bigg(4 ds^2(AdS_5) +\frac{ -y\partial_{y}D}{1- y \partial_{y}D} ds^2(S^2)+ \frac{-\partial_{y}D}{y} (dy^2+ e^{D} dr^2)\bigg)+4e^{-2\lambda- \frac{2}{3} \Phi+D} r^2 d\psi^2,\nn\\[2mm]
C_1&=-\frac{2y (2+r \partial_r D)}{ \partial_y D} e^{-4\lambda- \frac{4}{3} \Phi}d\psi,~~~ C_3=2 y^3 e^{-6\lambda}d\psi\wedge \text{Vol}(S^2),\nn\\[2mm]
 H&= \left(d\left[e^{-6\lambda}\frac{y^2}{\partial_y D}(2y \partial_y D+ r \partial_r D)\right]- \partial_y(e^{D}) r dr+ r \partial_r D dy\right)\wedge \text{Vol}(S^2)
\end{align}
where $F_2= dC_1$ and $F_4 = dC_3- H\wedge C_1$. The alert reader might notice that this solution does not actually correspond to imposing rotational symmetry in the $(x_1,x_2)$ plane of \cite[Eq.~(3.3)]{gaiotto-maldacena} then reducing to IIA along the isometry one has imposed on the solution. The reason for this, as we show in appendix \ref{appendix: AdS5}, is that reducing on this isometry breaks supersymmetry. The resolution is to perform the reduction on a linear combination of the two\footnote{We are ignoring the Cartan of the $S^2$ as reducing on this would break the SU(2) part of the R-Symmetry.}  U(1)'s at ones disposal in the axially symmetric AdS$_5$ M-theory solutions which results in \eqref{eq:Ads5sol}.

We thus have reproduced the solutions of \cite{lin-lunin-maldacena}, without any loss of generality. Thus, these solutions are the only one in massive IIA within our ``minimal'' class of manifolds. (This conclusion is similar to \cite{colgain-stefanski} in IIB.) If  $\mathcal{N}=2$ $AdS_5$ solutions with non-zero Romans mass do exist, R-symmetry must act in a more complicated way, such as with a squashed $S^3$ as discussed in section \ref{sec:ansatz}.

% subsection ads5 (end)

% section ex (end)

\section*{Acknowledgements}
We are supported in part by INFN and by the European Research Council under the European Union's Seventh Framework Program (FP/2007-2013) -- ERC Grant Agreement n. 307286 (XD-STRING). AT has also received support by the MIUR-FIRB grant RBFR10QS5J ``String Theory and Fundamental Interactions''.

\appendix

\section{Spinors on $S^2$} % (fold)
\label{app:s2}
To begin let us consider an arbitrary 2d Euclidean space. In general one can define the chiral spinors
\beq
\xi_{\pm}\ ,\qquad \xi^c_{\pm}= (\xi_{\pm})^c=B_2\xi_{\pm}^{*}=(\xi^c)_{\mp}\ .
\eeq
Let us work with the following basis of gamma matrices:
\beq
\gamma_1=\sigma_1 \ ,\qquad\gamma_2=\sigma_2 \ ,\qquad\hat{\gamma}=-i \gamma_1\gamma_2= \sigma_3 \ ,\qquad B_2=\sigma_2 \ ,\qquad B^{-1}_2\gamma_{\mu}B_2= -\gamma_{\mu}^{*}.
\eeq
In terms of the spinors we can then define the following scalars:
\beq\label{eq:scalars}
|\xi_{\pm}|^2= \alpha^2_{\pm} \ ,\qquad \bar{\xi}_-\xi_+=-(\xi^{\dag}_-\xi^c_+)^*=-\bar{\xi}_+\xi_-=(\xi^{\dag}_+\xi^c_-)^*= \beta,
\eeq
where $\alpha_{\pm}$ are real and $\beta$ is complex and as elsewhere $\overline{\xi}= (\xi^c)^{\dag}$. In general the scalars and spinors satisfy
\beq
\beta^*\xi_{\pm}=\pm \alpha^2_{\pm}\xi^c_{\mp} \ ,\qquad\beta \xi^c_{\pm}=\mp \alpha^2_{\pm}\xi_{\mp},
\eeq
We define the vectors
\beq \label{eq:vu}
v_{\pm}^{\mu}= \frac{1}{2} \xi_{\mp}^{\dag}\gamma^{\mu}\xi_{\pm}\ ,\qquad u_{\pm}^{\mu}= \frac{1}{2} \overline{\xi}_{\pm}\gamma^{\mu}\xi_{\pm}\ ,\qquad\overline{u}_{\pm}^{\mu}= \frac{1}{2} \xi^{\dag}_{\pm}\gamma^{\mu} \xi^c_{\pm},
\eeq
We then have the following bispinor identities:
\begin{equation}
\begin{split}
	\xi_{\pm}\otimes \xi^{\dag}_{\mp}&=-\xi^c_{\mp}\otimes \overline{\xi}_{\pm}= v_{\pm},\\
	\xi_{\pm}\otimes \overline{\xi}_{\pm}&= u_{\pm}\ ,\qquad
	\xi^c_{\pm}\otimes \xi^{\dag}_{\pm}= \overline{u}_{\pm},\\
	\xi_{\pm}\otimes \xi^{\dag}_{\pm}&=\left(\xi^c_{\pm}\otimes \overline{\xi}_{\pm}\right)^{*}=\frac{1}{2}\alpha_{\pm}^2 (1\pm\hat{\gamma}),\\
	\xi_{\pm}\otimes \overline{\xi}_{\mp}&=-\left(\xi^c_{\pm}\otimes \xi^{\dag}_{\mp}\right)^{*}=\pm\frac{1}{2}\beta (1\pm\hat{\gamma}) .
\end{split}	
\end{equation}
Using these we find that the vectors act on the spinors as
\begin{align}\label{eq:vector action}
&v_{\pm}\xi_{\mp}= \alpha^2_{\mp}\xi_{\pm}\ ,\qquad v_{\pm}\xi^c_{\pm}=\mp \beta^{*}\xi_{\pm},\nn\\[2mm]
&u_{\pm}\xi_{\mp}=\mp \beta\xi_{\pm}\ ,\qquad\overline{u}_{\pm}\xi_{\pm}= \alpha_{\pm}^2 \xi^c_{\pm}\ ,\qquad u_{\pm} \xi^c_{\pm}= \alpha^2_{\pm} \xi_{\pm}\ ,\qquad \overline{u}_{\pm} \xi^c_{\mp}=\pm\beta^{*}\xi^c_{\pm},
\end{align}
with all else giving zero. It is then rather simple to show that
\beq
v_{+} . v_{-} = \frac{1}{2} \alpha^2_+\alpha^2_-\ ,\qquad u_{\pm}.\overline{u}_{\pm} =\frac{1}{2} \alpha_{\pm}^4\ .
\eeq
This tells us that $v_{+}$ defines a complex vector whose conjugate is $v_-$, and $u_{\pm}$ define two complex vectors whose conjugates are $\overline{u}_{\pm}$.
For completeness we quote the remaining non vanishing inner products of the vectors:
\beq
u_{\pm}.v_{\mp}= \mp\frac{1}{2}\beta \alpha_{\pm}^2 \ ,\qquad \overline{u}_{\pm}. v_{\pm}=\mp\frac{1}{2}\beta^*\alpha_{\pm}^2 \ ,\qquad u_{\pm} . u_{\mp}=-\frac{1}{2}\beta^2 \ ,\qquad\overline{u}_{\pm} . \overline{u}_{\mp}=-\frac{1}{2}(\beta^*)^2,
\eeq
although they are not used in the main text. We will however need certain Lie brackets. For instance one can show that
\beq\label{eq: comutator}
[u_{\pm},\overline{u}_{\mp}]=\frac{1}{2}\gamma_{\mu\nu}\bigg(-4 v^{\mu}_{\pm}v^{\nu}_{\pm}+\overline{\xi}_{\pm}\gamma^{\mu\nu}v_{\pm}\xi^c_{\mp}\bigg)=0\ .
\eeq
The full list of non vanishing Lie brackets is
\begin{equation}
\begin{split}
	&[u_{\pm},\overline{u}_{\pm}] = \pm \alpha^4_{\pm}\hat\gamma \ ,\qquad~[u_{\pm},u_{\mp}]=\pm\beta^2\hat\gamma,\\
	&[v_{\pm},u_{\mp}]=\alpha^2_{\mp} \beta \hat\gamma \ ,\qquad[v_{\pm},\overline{u}_{\pm}]=-\alpha^2_{\pm} \beta^*\hat \gamma\ .	
\end{split}
\end{equation}

On $S^2$, there exist in particular Killing spinors, namely $\xi_\pm$ which satisfy
\beq\label{eq:KSES2}
\nabla_{\mu}\xi_{\pm}=\frac{i}{2}\gamma_{\mu}\xi_{\mp} \ ,\qquad\nabla_{\mu}\xi^c_{\pm}=\frac{i}{2}\gamma_{\mu}\xi^c_{\mp}\ .
\eeq	
From these it follows that
\beq
\nabla(\xi^{\dag}_{\pm}\xi_{\pm})=\nabla(\overline{\xi}_{\pm}\xi^c_{\pm})= \mp i(v_{+}-v_{-}) \ ,\qquad\nabla(\overline{\xi}_{\mp}\xi_{\pm})=\mp i(u_{+}-u_{-}) \ ,\qquad \nabla(\xi^{\dag}_{\mp}\xi^c_{\pm})=\mp i(\overline{u}_{+}-\overline{u}_{-}),
\eeq
which tell us that none of the scalar bilinears in \eqref{eq:scalars} are constant. However certain combinations are: for instance we have $
\nabla ( \alpha_{+}^2+\alpha_{-}^2) = 0$, which we can use to set
\beq\label{eq: choice}
\alpha_{+}^2+\alpha_{-}^2=1
\eeq
without loss of generality.

We know that $S^2$ has a global SU(2) isometry; thus we should be able to define three Killing vectors $K^a$ whose Lie brackets realize the SU(2) Lie algebra:
\begin{equation}
	[K^a,K^b]= \epsilon^{abc} K^c\ .
\end{equation}
This means that they should obey
\beq\label{eq:SU2}
\nabla_{\mu}K^a_{\nu}=\epsilon_{abc}K^b_{\mu}K^c_{\nu}\ .
\eeq
These Killing vectors can be found in terms of spinor bilinears as 
\begin{align}\label{eq:Killingvecs}
&(K^1)^{\mu}=(u^{\mu}_++  u^{\mu}_-+\overline{u}^{\mu}_++\overline{u}^{\mu}_-),\nn\\[2mm]
&(K^2)^{\mu} =-i (u^{\mu}_++  u^{\mu}_--(\overline{u}^{\mu}_++\overline{u}^{\mu}_-)) ,\\[2mm]
&(K^3)^{\mu} = 2(v^{\mu}_+ +v^{\mu}_-).\nn
\end{align}
To see that these indeed satisfy (\ref{eq:SU2}), we can compute their covariant derivatives using (\ref{eq:vu}):
\begin{align}\label{eq: Killing}
\nabla_{\mu} (K^1 + i K^2)_{\nu} &=-i (\overline{\xi}_+ \gamma_{\mu\nu}\xi_-+\overline{\xi}_- \gamma_{\mu\nu}\xi_+)= 2 \beta\epsilon_{\mu\nu},\nn\\[2mm]
\nabla_{\mu} (K^1 - i K^2)_{\nu} &=-i (\xi^{\dag}_+ \gamma_{\mu\nu}\xi^c_-+\xi^{\dag}_- \gamma_{\mu\nu}\xi^c_+)=2 \beta^*\epsilon_{\mu\nu},\nn\\[2mm]
\nabla_{\mu} (K^3)_{\nu} &=-i (\xi^{\dag}_+ \gamma_{\mu\nu}\xi_++\xi^{\dag}_- \gamma_{\mu\nu}\xi_-)= (\alpha_+^2-\alpha_-^2)\epsilon_{\mu\nu}\ .
\end{align}
These are all antisymmetric; so indeed $\nabla_{(\mu} K^a_{\nu)}=0$, and so the $K^a$ are Killing vectors. %Note by $\epsilon_{\mu\nu}$ we mean the curved space version. 
To show they obey the SU(2) relations, we can use the Lie bracket in \eqref{eq: comutator} and that
\beq
\frac{1}{2}\epsilon_{abc}[K^b,K^c]=\epsilon_{abc}K^b_{\mu}K^c_{\nu}\gamma^{\mu\nu} \ ,\qquad~ \hat\gamma = -\frac{i}{2}\epsilon_{\mu\nu}\gamma^{\mu\nu}\ .
\eeq
For example
\beq
\epsilon_{3bc}K^b_{\mu}K^c_{\nu}\gamma^{\mu\nu}=[K^1,K^2]=2i([u_+,\overline{u}_+]+[u_-,\overline{u}_-])=(\alpha^2_++\alpha^2_-)(\alpha^2_+-\alpha^2_-)\epsilon_{\mu\nu}\gamma^{\mu\nu}\ ,
\eeq
which given \eqref{eq: choice} is equal to $\nabla_{\mu} (K_3)_{\nu}\gamma^{\mu\nu}$. One can check the other SU(2) relations in a similar fashion. 

Having established that $K^a$ are the SU(2) Killing vectors, we choose to parameterise them as
\beq \label{eq:Ka}
K^a = \epsilon_{abc} y_b dy_c \ ,\qquad \ ,
\eeq
in terms of the ``embedding coordinates'' $y^a$ for the $S^2$, which obey
\begin{equation}\label{eq:ya}
	y_1^2+y_2^2+y_3^2=1\ .
\end{equation}
Consistency of the SU(2) algebra relation with \eqref{eq: Killing} then implies that
\beq\label{eq: y}
\alpha^2_{\pm}=\frac{1}{2}(1\pm y_3) \ ,\qquad\beta= \frac{1}{2} (y_1+i y_2)\ .
\eeq

The spinorial Lie derivative along one of these Killing vectors is given by
\begin{align}
\mathcal{L}_{K^a} \xi_{\pm} &= (K^a)^{\mu}\nabla_{\mu}\xi_{\pm}+\frac{1}{4} \nabla_{\mu}K^a_{\nu}\gamma^{\mu\nu}\xi_{\pm},\nn\\[2mm]
&=\frac{i}{2}\bigg( K^a \xi_{\mp}- \frac{i}{2}(\nabla K^a )\xi_{\pm}\bigg)
\end{align}
and similarly for $\xi^c_{\pm}$. Using \eqref{eq:vector action} and \eqref{eq: Killing} one finds
\begin{align}
&K^1\xi_{\pm} = \mp\beta \xi_{\mp} -\alpha^2_{\pm} \xi^c_{\pm} \ ,\qquad K^1\xi^c_{\pm}= \alpha^2_{\pm}\xi_{\pm}\mp \beta^* \xi^c_{\mp}\ , \qquad (\nabla K^1 )=2i(\beta+\beta^*)\hat\gamma\nn\\[2mm]
&K^2\xi_{\pm} = -i(\pm\beta \xi_{\mp} -\alpha^2_{\pm} \xi^c_{\pm})\ , \qquad K^2\xi^c_{\pm}=i( \alpha^2_{\pm}\xi_{\pm}\mp \beta^* \xi^c_{\mp})\ , \qquad (\nabla K^2 )=2(\beta-\beta^*)\hat\gamma\nn\\[2mm]
&K^3\xi_{\pm}= 2\alpha^2_{\pm}\xi_{\mp}\ , \qquad K^3\xi^c_{\pm}= -2\alpha^2_{\pm}\xi^c_{\mp}\ , \qquad\qquad (\nabla K^3 )=2i(\alpha^2_+-\alpha^2_-)\hat\gamma,
\end{align}
which is sufficient to show that 
\beq\label{eq:doubletS2}
\vec{\xi}_{\pm}=\left(\begin{array}{l}
\xi_{\pm}\\
\xi^c_{\mp} \end{array}\right)
\eeq
transforms as a doublet. For instance,
\beq
K^3 \vec{\xi}_{\mp}-\frac{i}{2} (\nabla K^3)\vec{\xi}_{\pm}=\left(\begin{array}{cc}
2\alpha^2_{\mp}\mp (\alpha_{\pm}^2-\alpha_{\mp}^2)\hat\gamma & 0 \\
0&-2\alpha^2_{\pm}\mp (\alpha_{\pm}^2-\alpha_{\mp}^2)\hat\gamma
\end{array}\right)\left(\begin{array}{c}
\xi_{\pm}\\
\xi^c_{\mp} \end{array}\right)=\sigma_3\vec{\xi}_{\pm}.
\eeq
Performing similar calculations for $K^{1,2}$ one can show that in general
\beq
\mathcal{L}_{K^a} \vec{\xi} = \frac{i}{2}\sigma_a\vec{\xi}\ .
\eeq
Actually \eqref{eq:doubletS2} is not the most general doublet: we can add a phase to each component without changing the transformation properties. In fact the most general doublet we can write is
\beq\label{eq:doubletS22}
\vec{\xi}_{\pm}=\left(\begin{array}{l}
\xi_{\pm}\\
e^{i \delta}\xi^c_{\mp} \end{array}\right)\ .
\eeq
A possible phase in the first entry can be absorbed into the definition of $\xi_{\pm}$.

% section s2 (end)

\section{Pure spinors from six to four dimensions} % (fold)
\label{app:psp4}

In terms of the vectors introduced in appendix \ref{app:s2}, one can parametrize the bispinors on $S^2$ as
\begin{align}
&\xi_{\pm}\otimes\xi^{\dag}_{\pm}=\frac{1}{4}(1\pm y_3)(1\mp i e^{2C}\text{Vol}(S^2))\ , \qquad \xi_{\pm}\otimes\xi^{\dag}_{\mp}= \frac{1}{4} e^{C}(K_3\pm i dy_3),\nn\\[2mm]
&\xi_{\pm}\otimes\overline{\xi}_{\pm}=\frac{1}{4} e^{C}(K_z\pm i dz)\ , \qquad \xi_{\pm}\otimes\overline{\xi}_{\mp}=\pm\frac{1}{4}z(1\mp i e^{2C}\text{Vol}(S^2)),
\end{align}
where we define $z=y_1+i y_2$, $K_z= K_1+ i K_2$ to ease presentation. This means the 6d bispinors implied by \eqref{eq: simp6dspinors} are
\begin{align}\label{eq: bi spin sim}
\Phi_-&=\frac{1}{4}\bigg[\Psi^1_+-\Psi^1_-+ y_3(\Psi^1_++\Psi^1_-)+ e^{C} K_3\wedge(\Psi^2_++\Psi^2_-)\\[2mm]
&+ i\bigg(e^{C} dy_3\wedge (\Psi^2_+-\Psi^2_-)- e^{2C} \text{Vol}(S^2)\wedge (\Psi^1_++\Psi^1_-)- e^{2C}y_3 \text{Vol}(S^2)\wedge (\Psi^1_+-\Psi^1_-)\bigg) \bigg],\nn\\[2mm]
\Phi_+&=\frac{1}{4}\bigg[z(\tilde\Psi^2_+-\tilde\Psi^2_-)- e^{C}K_z\wedge(\tilde\Psi^1_+-\tilde\Psi^1_-)- i \bigg(e^{C}dz\wedge(\tilde\Psi^1_++\tilde\Psi^1_-)+ z e^{2C}\text{Vol}(S^2)\wedge(\tilde\Psi^2_++\tilde\Psi^2_-)\bigg)\bigg].\nn
\end{align}
Here
\begin{align}
&\Psi^1_{\pm}= \eta^1_{\pm}\otimes\eta^{2\dag}_{\mp}\ , \qquad ~\Psi^2_{\pm}= \eta^1_{\pm}\otimes\eta^{2\dag}_{\pm},\nn\\[2mm]
&\tilde\Psi^1_{\pm}= \eta^1_{\pm}\otimes\overline{\eta}^{2}_{\mp}\ , \qquad ~\tilde\Psi^2_{\pm}= \eta^1_{\pm}\otimes\overline{\eta}^{2}_{\pm}.
\end{align}
These are in fact the even and odd parts of (\ref{eq:nochiralPSI}).

Plugging \eqref{eq: bi spin sim} in the supersymmetry conditions (\ref{eq:psp6d}) gives 8 independent conditions on $M_4$. To express these we decompose the NSNS three-form as
\beq \label{eq:HdB}
H= H_3 + e^{2C}H_1\wedge \text{Vol}(S^2)\ , \qquad  H= dB \ , \qquad  B= B_2 + e^{2C} B_0 \text{Vol}(S^2).
\eeq
We obtain two equations from (\ref{eq:Phi+}):
\begin{subequations}\label{eq: Phip}
	\begin{align}
	d_{H_3}\big(e^{3A+C-\Phi} \tilde{\Psi}^{\text{odd}}\big)&=i e^{3A-\Phi}\tilde{\Psi}_{\hat{\gamma}}^{\text{even}},\label{eq: Phip a}\\[2mm]
	d_{H_3}\big(e^{3A+2C-\Phi} \tilde{\Psi}^{\text{even}}\big)&=i e^{3A+2C-\Phi}H_1\wedge \tilde{\Psi}_{\hat{\gamma}}^{\text{even}}+2i e^{3A+C-\Phi} \tilde{\Psi}_{\hat{\gamma}}^{\text{odd}};\label{eq: Phip b}
	\end{align}
\end{subequations}
four from (\ref{eq:RePhi-}):
\begin{subequations}\label{eq: RePhim}
	\begin{align}
	d_{H_3}\big(e^{2A-\Phi}\text{Re}\Psi_{\hat\gamma}^{\text{odd}}\big)&=0,\label{eq: RePhim a}\\[2mm]
	d_{H_3}\big(e^{2A+C-\Phi}\text{Re}\Psi_{\hat\gamma}^{\text{even}}\big)&=e^{2A-\Phi} \text{Re}\Psi^{\text{odd}},\label{eq: RePhim b}\\[2mm]
	d_{H_3}\big(e^{2A+2C-\Phi}\text{Im}\Psi^{\text{odd}}\big)&=e^{2A+2C-\Phi}H_1\wedge\text{Re}\Psi_{\hat\gamma}^{\text{odd}}\label{eq: RePhim c},\\[2mm]
	d_{H_3}\big(e^{2A+2C-\Phi}\text{Im}\Psi_{\hat\gamma}^{\text{odd}}\big)&=e^{2A+2C-\Phi}H_1\wedge\text{Re}\Psi^{\text{odd}}+2 e^{2A+C-\Phi}\text{Re}\Psi^{\text{even}}\ ;\label{eq: RePhim d}
	\end{align}	
\end{subequations}
and finally, two from (\ref{eq:ImPhi-}):
\begin{subequations}\label{eq: ImPhim}
	\begin{align}
	d_{H_3}\big(e^{4A+C-\Phi}\text{Re}\Psi^{\text{even}}\big)&=e^{4A-\Phi}\text{Im}\Psi^{\text{odd}}\label{eq: ImPhim a},\\[2mm]
	d_{H_3}\big(e^{4A+2C-\Phi}\text{Re}\Psi_{\hat\gamma}^{\text{odd}}\big)&=e^{4A+2C-\Phi}H_1\wedge\text{Im}\Psi^{\text{odd}}-2e^{4A+C-\Phi}\text{Im}\Psi^{\text{even}}\ .\label{eq: ImPhim b},
	\end{align}	
\end{subequations}

The remaining conditions imply the fluxes. To express these we decompose the internal part of RR polyforms as
\beq
f =  F + e^{2C} \text{Vol}(S^2)\wedge G,
\eeq
which leads to the flux conditions
\begin{subequations}\label{eq:4dflux}
	\begin{align}
	d_{H_3}\big(e^{4A-\Phi}\text{Im}\Psi_{\hat\gamma}^{\text{odd}}\big)&=\frac{1}{2}e^{4A}\star_4\lambda(G),\label{eq:4dflux a}\\[2mm]
	d_{H_3}\big(e^{4A+2C-\Phi}\text{Re}\Psi^{\text{odd}}\big)&=-e^{4A+2C-\Phi}H_1\wedge\text{Im}\Psi_{\hat\gamma}^{\text{odd}}-\frac{1}{2}e^{4A+2C}\star_4\lambda(F)\label{eq:4dflux b}.
	\end{align}	
\end{subequations}
These conditions are rather restrictive: for example, it is already clear that the zero form parts of $\Psi$ and $\tilde\Psi_{\hat\gamma}$ must vanish, (\ref{eq:0f}). 

% section psp4 (end)

\section{Alternative classification of $b=a_1=0$ case}
\label{alt 4dsols}
In this appendix we give an alternative classification to section \ref{sub:4dsols}, in term of $B_0$ rather than $H_1= e^{-2C}d(e^{2C}B_0)$.
Upon setting $b=a_1=0,~a_2=1$ one can show that the supersymmetry conditions of appendix \ref{app:psp4} all follow from
\begin{subequations}
\begin{align}
&B_2 = 0,\label{alt4susy a}\\[2mm]
&d(e^{A}w)=0,~~~d(e^{2A+C-\Phi})+ e^{2A-\Phi}v_2=0,~~~d(e^{-2A+\Phi}(v_1+ B_0 v_2))=0,\label{alt4susy b}\\[2mm]
&d(e^{-\Phi} v_1)\wedge w\wedge \overline{w}=0,~~~d(e^{2C-\Phi} (B_0v_1-v_2))\wedge w\wedge \overline{w}=0,\label{alt4susy c}
\end{align}
\end{subequations}
One can solve the conditions \eqref{alt4susy b} locally by defining the vielbein in terms of local coordinates $(x_1,...,x_4)$ as
\beq
v_1= e^{2A-\Phi} dx_1+ B_0 e^{-2A+\Phi} dx_2,\qquad v_2=- e^{-2A+\Phi}dx_2,\qquad  w= e^{-A}(dx_3+i dx_4),\qquad  x_2 = e^{2A+B-\Phi},\nn
\eeq
it then follows that \eqref{alt4susy c} imposes the PDEs
\begin{align}\label{eq: BPS ba10}
&\partial_{x_2}\left(e^{2A-2\Phi}\right)=\partial_{x_1}\left(e^{-2A}B_0\right),\nn\\[2mm]
&\partial_{x_2}\left(x_2^2e^{-2A}B_0\right)= \partial_{x_1}\left(x_2^2e^{-6A+2\Phi}(1+ B_0^2)\right),
\end{align}
but place no restriction on the various functions dependence on $(x_3,x_4)$, so we only have an a prior $\mathbb{R}_{1,3}$ factor in these solutions.
Following the same prescription as the main text on can establish that the fluxes are given by
\begin{align}
B=&x_2^2 e^{-4A+2\Phi}B_0\text{Vol}(S^2),\nn\\[2mm]
F_2=&\big(\partial_{x_4}(e^{2A-2\Phi})dx_3-\partial_{x_3}(e^{2A-2\Phi})dx_4\big)\wedge dx_1+\big(\partial_{x_4}(e^{-2A}B_0)dx_3-\partial_{x_3}(e^{-2A}B_0)dx_4\big)\wedge dx_2\nn\\[2mm]
&-\partial_{x_1}(e^{-4A})dx_3\wedge dx_4,\nn\\[2mm]
F_4=&B\wedge F_2+x_2^2\bigg[-\big(\partial_{x_4}(e^{-2A}B_0)dx_3-\partial_{x_3}(e^{-2A}B_0)dx_4\big)\wedge dx_1\\[2mm]
&+\big(\partial_{x_4}(e^{-6A+2\Phi}(1+B_0^2))dx_3-\partial_{x_3}(e^{-6A+2\Phi}(1+B_0^2))dx_4\big)\wedge dx_2 -\partial_{x_2}(e^{-4A})dx_3\wedge dx_4 \bigg]\wedge\text{Vol}(S^2),\nn
\end{align}
the metric is 
\beq
ds^2= e^{2A}ds^2(\mathbb{R}_{1,3}) + e^{-4A+2\Phi}\bigg(dx_2^2+ x_2^2 ds^2(S^2)\bigg)+ e^{-2A}\bigg(dx_3^2+ dx_4^2\bigg)+  e^{4A-2\Phi}\bigg(dx_1+ B_0 e^{-4A+2\Phi}dx_2\bigg)^2\nn.
\eeq
Ensuring that the fluxes obey the correct Bianchi identities imposed the following PDEs
\begin{align}\label{eq: 4dbianchis}
&\partial^2_{x_3} (e^{2A-2\Phi})+\partial^2_{x_4} (e^{2A-2\Phi})+\partial^2_{x_1}(e^{-4A})=0,\nn\\[2mm]
&\partial^2_{x_3} (e^{-2A}B_0)+\partial^2_{x_4} (e^{-2A}B_0)+\partial_{x_1} \partial_{x_2}(e^{-4A})=0,\nn\\[2mm]
&\partial^2_{x_3}(x_2^2 e^{-6A+2\Phi}(1+B_0^2))+\partial^2_{x_4}(x_2^2 e^{-6A+2\Phi}(1+B_0^2))+ \partial_{x_2}(x_2^2\partial_{x_2}(e^{-4A}))=0.
\end{align}

We can take the first equation of \eqref{eq: BPS ba10}  as an integrability condition implying the existence of $h(x_1,x_2,x_3,x_4)$ such that
\beq
e^{2A-2\Phi}= \partial_{x_1} h,\qquad ~ e^{-2A}B_0= \partial_{x_2} h,
\eeq
the second equations then implies
\beq
\frac{1}{x_2^2}\partial_{x_2}(x_2^2 \partial_{x_2} h)= \partial_{x_1}k,\qquad  k=\frac{e^{-4A}+ (\partial_{x_2} h)^2}{\partial_{x_1}h},
\eeq
and the Bianchi identity conditions become
\begin{align}
\partial_{x_i}\big(\partial^2_{x_3}h+\partial^2_{x_4}h+ \partial_{x_1}(e^{-4A})\big)&=0,\qquad  i=1,2,\nn\\[2mm]
\partial^2_{x_3}k+\partial^2_{x_4}k+ \frac{1}{x_2^2}\partial_{x_2}(x_2^2\partial_{x_2}(e^{-4A}))&=0.
\end{align}
The first of these can be integrated in terms of an arbitrary function $l(x_3,x_4)$ to give the coupled system
\begin{align}
&k=\frac{e^{-4A}+ (\partial_{x_2} h)^2}{\partial_{x_1}h},\nn\\[2mm]
&\frac{1}{x_2^2}\partial_{x_2}(x_2^2 \partial_{x_2} h)= \partial_{x_1}k,\nn\\[2mm]
&\partial^2_{x_3}h+\partial^2_{x_4}h+ \partial_{x_1}(e^{-4A}) = l,\nn\\[2mm]
&\partial^2_{x_3}k+\partial^2_{x_4}k+ \frac{1}{x_2^2}\partial_{x_2}(x_2^2\partial_{x_2}(e^{-4A}))=0,
\end{align}
which looks hard to disentangle.

\section{The General $AdS_5\times S^2\times \mathcal{M}_4$ solutions in M-theory}\label{appendix: AdS5}
In  \cite{lin-lunin-maldacena} a class of  $AdS_5\times S^2\times \mathcal{M}_4$ solution in M-theory was presented which was later argued to give the general local form of such solutions \cite{gaiotto-maldacena,OColgain:2010ev}. They take the form
\begin{align}\label{eq: LLM}
ds^2&=e^{2\lambda}\bigg(4 ds^2(AdS_5) + y^2 e^{-6\lambda} ds^2(S^2)\bigg)+ \frac{4}{(1-y \partial_yD)}e^{2\lambda}\left(d\chi + V\right)^2,\nn\\[2mm]
&+ \frac{-\partial_y D}{y}e^{2\lambda}\bigg(dy^2+ e^{D} (d\hat x_1^2+ d\hat x_2^2)\bigg),\\[2mm] 
e^{-6\lambda}&=\frac{- \partial_yD}{y(1- y \partial_yD)},~~~V= \frac{1}{2}(\partial_{\hat \hat x_2}D d \hat x_1- \partial_{\hat x_1}Dd \hat x_2),\nn\\[2mm]
G_4&= \bigg[2(d\chi+ V)\wedge d\big( y^3 e^{-6\lambda}\big)+ 2 y(1-y^2 e^{-6\lambda}) dV+\partial_{y} D \hat dx_1\wedge \hat dx_2\bigg]\wedge \text{Vol}(S^2)\nn,
\end{align}
and are governed by the Toda equation
\beq
\partial_{\hat x_1}^2 D+ \partial_{\hat x_2}^2 D+ \partial_{y}^2 e^{D}=0.
\eeq
Starting from \cite[App.~B]{lunin}, Wick rotating\footnote{One could also start from \cite{bah}, and avoid this step.} and using the formula in \cite{gauntlett-martelli-sparks-waldram-M} relating SU(2) to SU(3) structures for $AdS_5$ solutions of M-theory we can express the SU(3) structure of \eqref{eq: LLM} as
\begin{align}
K&= e^{-2\lambda}\bigg( y_3( dy+2 y d\rho)+ y dy_3\bigg),\nn\\[2mm]
J&=\frac{i}{2}\bigg(\hat{E}_1\wedge \overline{\hat{E}}_1+\hat{E}_2\wedge \overline{\hat{E}}_2+\hat{E}_3\wedge \overline{\hat{E}}_3\bigg),~~~
\Omega= \hat{E}_1\wedge \hat{E}_2\wedge \hat{E}_3,
\end{align}
where
\begin{align}\label{eq:GMniceframe}
\hat{E}_1&= \sqrt{\frac{-\partial_yD}{y}}e^{\lambda+\frac{1}{2}D}\bigg(d\hat x_1+ i d\hat x_2\bigg),\nn\\[2mm]
\hat{E}_2&= e^{-2\lambda}\bigg((y_1+i y_2) (dy+2yd\rho)+yd(y_1+i y_2) \bigg),\nn\\[2mm]
\hat{E}_3&= -e^{i \chi}\frac{2}{\sqrt{1- y \partial_yD}}e^{-2\lambda}\bigg(d\rho+\frac{1}{2}\partial_y Ddy + i (d\chi+ V)\bigg),
\end{align}
and we parametrises $ds^2(AdS_5)= e^{2\rho} ds^2(\mathbb{R}_{1,3})+ d\rho^2$ and $ds^2(S^2)= dy_1^2+dy_2^2+dy_3^2$ for $y_1^2+y_2^2+y_3^2=1$. These G-structure forms obey the relations
\beq
d(e^{2(\rho+\lambda)} K) = d(e^{3(\rho+\lambda)}\Omega)=d(e^{2(\rho+ \lambda)}J\wedge J)+2e^{2(\rho+ \lambda)}G\wedge K=0,~~~ d(e^{4(\rho+ \lambda)}J)= e^{4(\rho+ \lambda)}\star_7G_4,\nn
\eeq
where $\star_7$ is taken on the part of the metric that is not $\mathbb{R}_{1,3}$, which means the solution is indeed supersymmetric \cite{kmt}.

We want to reduce this to IIA and see if it falls within the class of solutions in section \ref{sub:ads5}. To this end we  set $\hat x_1=r \cos\beta,~ \hat x_2= r\sin\beta$ and impose that $\beta$ is an isometry.
This modifies 
\beq
V= -\frac{1}{2}\partial_r D d\beta , ~~~ \hat{E}_1= e^{i\beta}\sqrt{\frac{-\partial_yD}{y}}e^{\lambda+\frac{1}{2}D}(dr+ i rd \beta),~~~d \hat x_1\wedge d\hat x_2= rdr\wedge d\beta,~~~ \frac{1}{r} \partial_{r} (r \partial_r D)+ \partial_y^2 e^{D}=0,\nn
\eeq
and leave the rest unchanged. We notice then that $\Omega$ now depends on the phase $e^{i(\chi+\beta)}$ which suggests that the U(1) factor of the total $U(1)\times SU(2)$ R-symmetry is in fact given by $\psi= \chi+\beta$, not $\chi$. In other words one needs to set $\chi=\psi-\beta$, before reducing on $\beta$ if we want supersymmetry to be preserved. The flux then decomposes as
\begin{align}
G_4 &= dC_3 + e^{2C}H_1\wedge \text{Vol}(S^2)\wedge d\beta,~~~C_3=2y^3 e^{-6\lambda}d\psi\wedge \text{Vol}(S^2)\nn\\[2mm]
e^{2C}H_1&= d\bigg[e^{-6\lambda}\frac{y^2}{\partial_y D}( 2y \partial_y D+ r \partial_r D)\bigg]-\partial_y(e^{D}) rdr+ r\partial_r Ddy.
\end{align}
Substituting for $\chi$ in eq \eqref{eq:GMniceframe} and rotating to a reduction frame we arrive at
\begin{align}
K&= e^{-\frac{1}{3}\Phi}\bigg( e^{-2A+\Phi} 4 y e^{2\rho}dy_3+ y_3e^{-2A+\Phi}d(4 ye^{2\rho})\bigg),\nn\\[2mm]
\hat{E_1}&= -e^{-\frac{1}{3}\Phi}  e^{-A}d\bigg( 4r e^{\rho+\frac{1}{2} D}e^{i\psi}\bigg),\nn\\[2mm]
\hat{E_2}&=-e^{-\frac{1}{3}\Phi}\bigg( e^{-2A+\Phi}4ye^{2\rho} d(y_1+ i y_2)+ (y_1+i y_2)e^{-2A+\Phi}d(4y e^{2\rho})\bigg),\nn\\[2mm]
\hat{E_3}&= -e^{-\frac{1}{3}\Phi}e^{2A-\Phi}\bigg(d\left(\frac{e^{-2\rho}}{-4 \partial_y D}y(2+r \partial_rD)\right)+ \frac{1}{4y e^{2\rho}}e^{2C}H_1\bigg)\nn\\[2mm]
&+ i e^{\frac{2}{3}\Phi}\bigg(d\beta- 2 y \frac{(2+ r\partial_r D)}{\partial_y D} e^{-4(A-\rho)}d\psi\bigg),\nn\\[2mm]
e^{2A}&= 4e^{2\rho+2\lambda+\frac{2\Phi}{3}},~~~ e^{4\Phi}= \frac{e^{-4\lambda}}{-y\partial_{y}D}\bigg[r^2 (\partial_yD)^2 e^{D}+ y^2(2+ r \partial_rD)^2 \bigg],
\end{align}
from which we read off the local IIA coordinates of section \ref{sub:ads5}
\beq
x_1=e^{-2\rho}\frac{2+r \partial_rD}{4(1- y\partial_y D)},~~~ x_2=4y e^{2\rho},~~~ x_3+i x_4 =  4r e^{\rho+\frac{1}{2} D}e^{i\psi}.
\eeq
by comparing to eq \eqref{eq:canonical6d vielbein}.

\bibliography{at}
\bibliographystyle{at}

\end{document}